\documentclass[aps,physrev,reprint,nofootinbib,superscriptaddress,longbibliography]{revtex4-2}
\usepackage[T1]{fontenc}
\usepackage{amsmath,amssymb,amsfonts,mathtools,bm}
\usepackage{graphicx}
\usepackage[colorlinks=true,allcolors=blue]{hyperref}
\usepackage{scalerel,tikz}
\usetikzlibrary{svg.path}
\definecolor{orcidlogocol}{HTML}{A6CE39}
\tikzset{orcidlogo/.pic={
    \fill[orcidlogocol]
    svg{M256,128c0,70.7-57.3,128-128,128C57.3,256,0,198.7,0,128C0,57.3,57.3,0,128,0C198.7,0,256,57.3,256,128z};
    \fill[white]
    svg{M86.3,186.2H70.9V79.1h15.4v48.4V186.2z} svg{M108.9,79.1h41.6c39.6,0,57,28.3,57,53.6c0,27.5-21.5,53.6-56.8,53.6h-41.8V79.1zM124.3,172.4h24.5c34.9,0,42.9-26.5,42.9-39.7c0-21.5-13.7-39.7-43.7-39.7h-23.7V172.4z}
    svg{M88.7,56.8c0,5.5-4.5,10.1-10.1,10.1c-5.6,0-10.1-4.6-10.1-10.1c0-5.6,4.5-10.1,10.1-10.1C84.2,46.7,88.7,51.3,88.7,56.8z};
}}
\newcommand\orcidicon[1]{
    \href{https://orcid.org/#1}{\mbox{\scalerel*{
        \begin{tikzpicture}[yscale=-1,transform shape]
            \pic{orcidlogo};
        \end{tikzpicture}
    }{|}}}
}
\newcommand{\ltab}[1]{\parbox[t]{\linewidth}{\raggedright #1}}

\begin{document}

\title{Compressible Navier--Stokes Flow in Schr\"odinger-Type Variables}

\author{James R. Beattie\,\orcidicon{0000-0001-9199-7771}}
    \email{jbeattie@cita.utoronto.ca}
    \affiliation{Canadian Institute for Theoretical Astrophysics, 60 St. George Street, Toronto, ON M5S 3H8, Canada}
    \affiliation{Department of Astrophysical Sciences, Princeton University, Princeton, 08540, NJ, USA}
\author{Max Sokolova\,\orcidicon{0009-0006-8963-1635}}
    \affiliation{Department of Physics, University of Toronto, 60 St. George Street, Toronto, ON M5S 1A7, Canada}
\author{Khush Negandhi\,\orcidicon{0009-0007-5167-3669}}
    \affiliation{Department of Physics, University of Toronto, 60 St. George Street, Toronto, ON M5S 1A7, Canada}
\author{Bart Ripperda\,\orcidicon{0000-0002-7301-3908}}
    \affiliation{Canadian Institute for Theoretical Astrophysics, 60 St. George Street, Toronto, ON M5S 3H8, Canada}    \affiliation{Department of Physics, University of Toronto, 60 St. George Street, Toronto, ON M5S 1A7, Canada}
    \affiliation{D. A. Dunlap Department of Astronomy, University of Toronto, Toronto, ON M5S 3H4, Canada}
    \affiliation{Perimeter Institute for Theoretical Physics, Waterloo, ON N2L 2Y5, Canada}
\date{\today}

\begin{abstract}
    Fluid equations are nonlinear, dissipative, and non-Hamiltonian, which makes their relation to Schr\"odinger evolution and quantum algorithms nontrivial. We derive an exact Eulerian Cole--Hopf-type reformulation of isothermal compressible Navier--Stokes (NS) flow in Schr\"odinger-type amplitude variables. To our knowledge, this gives the first exact Cole--Hopf-type Schr\"odinger-variable reformulation of compressible NS flow. In two dimensions, a Helmholtz decomposition separates the velocity into compressive and vortical potentials, whose logarithmic transforms yield two scalar imaginary-time Schr\"odinger-type equations with nonlinear self-consistent potentials. We show that the mixed density-compressive amplitude \(\Psi_\alpha=\rho^\alpha\Theta^{1-2\alpha}\), where \(\rho\) is the density, \(\Theta\) is the compressive amplitude, and \(\alpha\neq 0,\,1/2\), satisfies a nonlinear Schr\"odinger-type equation with a vector-potential-coupled Laplacian. The transformed system is exactly equivalent to compressible NS and is nonlocal only through Helmholtz and Poisson projections. In three dimensions, the density-carrying equation retains the same vector-potential-coupled structure, while the solenoidal sector admits a compressible analogue of Ohkitani's \cite{Ohkitani2015_PRE} incompressible NS Cole--Hopf formulation. Unlike unitary hydrodynamic Schr\"odinger-flow representations, the present equations are imaginary-time heat or drift-diffusion equations with self-consistent potentials, but they remain an exact change of variables for compressible NS. A two-dimensional Kelvin--Helmholtz shear-layer calculation verifies the transformed equations against a direct compressible NS simulation. The formulation exposes operator structures that may be useful for reduced flow descriptions, quantum algorithms for operator evolution, and quantum partial differential equation solvers.
\end{abstract}


\maketitle

\section{Introduction}
\label{sec:introduction}
    The classical Cole--Hopf transformation linearizes the one-dimensional viscous Burgers equation by writing \(u(x,t)=-2\nu\,\partial_x\ln\theta(x,t)\), where \(u\) is the flow velocity, \(x\) is position, \(t\) is time, \(\nu\) is the viscosity, and \(\theta\) is a scalar amplitude \cite{Burgers1948,Hopf1950,Cole1951}. With this substitution, the nonlinear advection term \(u\,\partial_x u\) is absorbed into the heat equation, equivalently an imaginary-time Schr\"odinger-type equation, \(\partial_t\theta=\nu\,\partial_x^2\theta\), for the transformed variable. This elegant transformation is special, since it relies on potential flow and does not extend directly to incompressible Navier--Stokes (NS) flow, where the nonlinear advection term \(\bm{u}\cdot\bm{\nabla}\bm{u}\) is coupled to the pressure force \(-\bm{\nabla}p\). The pressure enforces the incompressibility constraint \(\bm{\nabla}\cdot\bm{u}=0\) through a nonlocal projection onto solenoidal, or divergence-free, velocity fields. Ohkitani \cite{Ohkitani2008} emphasized this nonlocal structure through alternative streamfunction and vorticity-based formulations, and later derived dynamical equations for streamfunctions, vector potentials, and velocity potentials \cite{Ohkitani2015_PRE}. These potential formulations provide natural variables for Cole--Hopf-type logarithmic transforms, even though they do not linearize the full incompressible equations \cite{Ohkitani2017}. They also suggest a route for extending the construction to compressible NS flow by reformulating the dynamics in terms of wavefunction-like amplitude variables with heat- and Schr\"odinger-type structure.

    For incompressible NS flow, Ohkitani \cite{Ohkitani2017} showed that a streamfunction \(\psi\) in two dimensions, or a vector streamfunction \(\bm{\psi}\) in three dimensions with \(\bm{u}=\bm{\nabla}\times\bm{\psi}\), can be transformed logarithmically into heat equations with nonlinear self-consistent potential terms, or equivalently, imaginary-time nonlinear Schr\"odinger equations. The resulting Feynman--Kac representation gives a regularity criterion and makes the self-consistent potential a diagnostic for localized structures with large vorticity gradients in two-dimensional turbulence. Vanon and Ohkitani \cite{VanonOhkitani2018} applied the same component-wise transform to three-dimensional incompressible NS simulations, where the potential highlights vortex reconnection and regions of intense vorticity interaction; they also derived an analogous construction for incompressible magnetohydrodynamics (MHD). In this study we determine whether this framework can be extended from incompressible to \emph{compressible} isothermal NS flow, broadening the applicability of these model reformulations to the ubiquitous, highly-compressible flows across the Universe \cite{2025NatAs...9.1195B,2025ApJ...982L..45B} and in Appendix~\ref{app:MHD} we extend it to compressible MHD\footnote{The extension of the non-ideal magnetic induction equation, \(\partial_t \bm{B} = \bm{\nabla}\times(\bm{u}\times\bm{B}) + \eta\Delta\bm{B}\), where \(\eta\) is the Ohmic resistivity coefficient, is structurally direct. After introducing a magnetic vector potential in Coulomb gauge, the component-wise logarithmic transform gives the same imaginary-time nonlinear Schr\"odinger-type form as the solenoidal velocity-potential transform in Sec.~\ref{sec:three_dimensional_ns}. The momentum equation is modified separately by the Lorentz force.}.

    The compressible case changes the problem in two essential ways. First, the velocity \(\bm{u}\) has both compressive and solenoidal components. In two dimensions we write \(\bm{u}=\bm{\nabla}\phi+\bm{\nabla}^{\perp}\psi\), where \(\phi\) is the compressive potential, \(\psi\) is the streamfunction, and \(\bm{\nabla}^{\perp}f\equiv(\partial_y f,-\partial_x f)\) is the skew gradient. Thus a single solenoidal potential is no longer sufficient. Second, the density \(\rho\) obeys a first-order continuity equation rather than a diffusion equation. In Appendix~\ref{app:obstruction}, we show that this imposes a restriction on direct transformations of \(\rho\) into a Schr\"odinger equation for many choices of \(f(\rho)\). Following Ohkitani's incompressible Cole--Hopf analogue \cite{Ohkitani2017}, we apply logarithmic transforms to both velocity potentials, \(\phi=-2\mu_c\ln\Theta\) and \(\psi=2\mu_s\ln\Xi\), where \(\Theta\) and \(\Xi\) are the compressive and solenoidal amplitudes and \(\mu_c\) and \(\mu_s\) set their transform scales. To incorporate density, we introduce the mixed amplitude \(\Psi_\alpha=\rho^\alpha\Theta^{1-2\alpha}\), with \(\alpha\neq 0,1/2\), whose powers are chosen so that the compressive Laplacian terms cancel. Our result is an exact Eulerian reformulation in terms of \(\Theta\), \(\Xi\), and \(\Psi_\alpha\), coupled nonlocally through Helmholtz and Poisson projections in two and three dimensions. The \(\Theta\) and \(\Xi\) sectors retain heat or imaginary-time Schr\"odinger structure, while \(\Psi_\alpha\) satisfies a nonlinear Schr\"odinger-type equation with a vector-potential-coupled Laplacian.

    A second motivation comes from quantum and quantum-inspired approaches to nonlinear partial differential equations (PDEs). Because quantum operations are linear and quantum measurement returns limited classical information, nonlinear field equations are often recast before either evolution or readout, i.e., before extracting classical observables or statistics from the encoded quantum state. Existing quantum-fluid work includes algorithms for targeted fluid benchmarks and nonlinear dissipative systems \cite{Gaitan2020_quantum_ns,Liu2021_dissipative_nonlinear_quantum,Oz2022_burgers_quantum,Pfeffer2022_quantum_reservoir_convection}, quantum lattice-Boltzmann algorithms for streamfunction-vorticity formulations of incompressible NS flow \cite{Budinski2022_quantum_nse}, hybrid variational methods for nonlinear Schr\"odinger--Poisson dynamics \cite{MoczSzasz2021_quantum_cosmo}, Madelung-type hydrodynamic reformulations, including geometric studies of the Madelung transform and quantum-walk implementations \cite{Madelung1927,KhesinMisiolekModin2018_geometric_madelung,Zylberman2022_madelung_hydro}, and quantum-inspired matrix-product-state compression of Vlasov--Poisson dynamics \cite{YeLoureiro2022_mps_vlasov}. Particularly relevant to the present paper, Meng and Yang \cite{MengYang2023_hydrodynamic_schrodinger} derive a hydrodynamic Schr\"odinger equation (HSE) by generalizing the Madelung transform to a two-component wave function for compressible or incompressible flows with finite vorticity and dissipation. Their quantum algorithm and turbulence tests focus on the incompressible Schr\"odinger flow, which resembles viscous turbulence in some structures and statistics but is not an exact NS reformulation. Closest in algebraic strategy, Uchida et al. \cite{Uchida2024_Burgers_quantum} use the Cole--Hopf transformation to map Burgers turbulence to a heat equation for the transformed field, and then extract selected velocity statistics from the quantum state under a perturbative readout approximation. Complementary hardware studies show that central moments and structure functions of amplitude-encoded velocity fields can be measured without full state tomography \cite{Goldack2026_velocity_stats}. At the same time, detailed complexity comparisons for the heat equation show that quantum advantage is task-dependent and can disappear once state preparation, error scaling, readout, and the best classical algorithms are included \cite{Linden2022_heat_equation}. Thus the present transformation should not be read as a quantum speedup claim. Rather, we rewrite the compressible isothermal NS (and MHD in Appendix~\ref{app:MHD}) equations as a coupled set of amplitude equations whose heat- and Schr\"odinger-type linear parts may be able to benefit from quantum algorithms for operator evolution, while leaving nonlinear potentials, elliptic projections, state preparation, and observable extraction as explicit challenges.\\
    \newline
    The gain is not that the compressible equations become linear. Rather, the transformation gives an exact operator decomposition of the flow. It separates compressive, vortical, and density-carrying degrees of freedom; identifies the scalar and vector self-consistent potentials through which nonlinear advection, pressure, and projection effects enter; and places the leading linear pieces in heat or drift-diffusion form. This structure makes the nonlinear couplings explicit enough to compare with incompressible Cole--Hopf analogues, to define possible reduced operators, and to isolate which terms would have to be retained or approximated in any quantum or quantum-inspired implementation.\\
    \newline
    The study is organized as follows. In Sec.~\ref{sec:two_dimensional_ns}, we develop the two-dimensional construction, beginning from the standard compressible NS equations and ending with a closed set of transformed amplitude equations. In Sec.~\ref{sec:three_dimensional_ns}, we give the three-dimensional analogue, including the vector-streamfunction formulation of the solenoidal sector. In Sec.~\ref{sec:numerical_validation}, we use a direct Kelvin--Helmholtz comparison as a consistency check on the exact transformed equations. In Sec.~\ref{sec:discussion}, we discuss the limitations, extensions, reduced-model possibilities, and quantum-computing relevance of the reformulation. In Appendix~\ref{app:green_operator}, we relate our Leray-projector formulation to the Green's function operator used by Ohkitani \cite{Ohkitani2015_PRE}; in Appendix~\ref{app:MHD}, we derive a compressible MHD extension; in Appendix~\ref{app:obstruction}, we make the restriction on pure density-only transforms explicit. In Appendix~\ref{app:other_transforms}, we show that several broader transform classes do not give a single closed second-order density equation.

\section{Two-Dimensional Navier--Stokes}
\label{sec:two_dimensional_ns}
\subsection{Governing equations}
    We consider the two-dimensional isothermal \((p=c_s^2\rho)\) compressible NS equations
    \begin{align}
        \partial_t \rho + \bm{\nabla}\cdot(\rho \bm{u}) &= 0,
        \label{eq:continuity_rho}
        \\
        \partial_t \bm{u} + \bm{u}\cdot\bm{\nabla} \bm{u}
        &=
        - c_s^2 \bm{\nabla} \ln \rho
        + \nu \Delta \bm{u}
        + (\zeta+\nu)\bm{\nabla}(\bm{\nabla}\cdot \bm{u}),
        \label{eq:momentum_u}
    \end{align}
    where \(c_s\) is the isothermal sound speed, \(\nu\) is the shear viscosity, and \(\zeta\) is the kinematic coefficient multiplying the additional compressive viscous term. The isothermal pressure term already appears as a gradient of \(\ln\rho\), so it is useful to make the logarithmic density a primary scalar variable, $s \equiv \ln \rho$. With this substitution the continuity equation becomes a transport equation with source \(-\bm{\nabla}\cdot\bm{u}\). This source depends only on the compressive component of the velocity, which we represent by the scalar potential introduced below.

    Throughout the two-dimensional derivation we assume either periodic boundary conditions, with zero-mean potentials used to fix the inverse Laplacian, or sufficiently decaying fields on \(\mathbb{R}^2\). The Helmholtz potentials are therefore understood modulo the usual constant freedoms. Nonlocal scalar potentials are fixed by the corresponding Poisson equations together with these conventions. The resulting zero-mode choices do not affect the reconstructed velocity field, but they fix the amplitudes used below.

\subsection{Helmholtz decomposition}
    In two dimensions we write the velocity as
    \begin{equation}
        \bm{u} = \bm{\nabla} \phi + \bm{\nabla}^\perp \psi,
        \qquad
        \bm{\nabla}^\perp \psi \equiv (\partial_y \psi,\,-\partial_x \psi).
    \end{equation}
    Then
    \begin{equation}
        \bm{\nabla}\cdot\bm{u} = \Delta \phi,
        \qquad
        \omega \equiv \partial_x u_y - \partial_y u_x = -\Delta \psi.
    \end{equation}
    In terms of \(s\), the continuity equation becomes
    \begin{equation}
        \partial_t s + \bm{u}\cdot\bm{\nabla} s = -\bm{\nabla}\cdot \bm{u} = -\Delta \phi, \label{eq:continuity_s_general}
    \end{equation}

\subsection{Potential formulation of the momentum equation}
    Using the two-dimensional identity
    \begin{equation}
        \bm{u}\cdot\bm{\nabla} \bm{u}
        =
        \bm{\nabla}\!\left(\frac{|\bm{u}|^2}{2}\right)
        +
        \bm{u}^\perp \,\omega,
        \qquad
        \bm{u}^\perp \equiv (-u_y,u_x),
    \end{equation}
    we rewrite \eqref{eq:momentum_u} as
    \begin{equation}
        \partial_t \bm{u}
        +
        \bm{\nabla}\!\left(\frac{|\bm{u}|^2}{2} + c_s^2 s\right)
        +
        \bm{u}^\perp \omega
        =
        \nu \Delta \bm{u}
        +
        (\zeta+\nu)\bm{\nabla}(\bm{\nabla}\cdot\bm{u}).
    \end{equation}
    We define
    \begin{equation}
        \bm{F} \equiv \bm{u}^\perp \omega,
    \end{equation}
    and decompose \(\bm{F}\) into gradient and skew-gradient parts,
    \begin{equation}
        \bm{F} = \bm{\nabla} \Phi_F + \bm{\nabla}^\perp \Psi_F,
    \end{equation}
    where
    \begin{align}
         \bm{\nabla}\cdot \bm{F} &= \Delta \Phi_F,
        \\
        \bm{\nabla}^\perp \cdot \bm{F} &=-\Delta \Psi_F, \\
        \bm{\nabla}^\perp\cdot\bm{F}
        &\equiv \partial_x F_2-\partial_y F_1.
    \end{align}
    Projecting the momentum equation onto the gradient and skew-gradient sectors gives
    \begin{align}
        \partial_t \phi + \frac{|\bm{u}|^2}{2} + c_s^2 s + \Phi_F
        &= (\zeta+2\nu)\Delta\phi,
        \label{eq:phi_eq}
        \\
        \partial_t \psi + \Psi_F
        &= \nu \Delta \psi.
        \label{eq:psi_eq}
    \end{align}

\subsection{Logarithmic variables}
    We now introduce logarithmic amplitudes for the compressive and vortical potentials,
    \begin{equation}
        \phi = -2\mu_c \tau,
        \qquad
        \psi = 2\mu_s \chi,
        \qquad
        \tau \equiv \ln \Theta,
        \qquad
        \chi \equiv \ln \Xi,
    \end{equation}
    where
    \begin{equation}
        \mu_c \equiv \zeta + 2\nu,
        \qquad
        \mu_s \quad \text{is the solenoidal transform scale.}
    \end{equation}
    One may specialize to the canonical choice \(\mu_s=\nu\), but we keep the symbols distinct to avoid conflating the transform normalization with the physical viscosity. Thus \(\mu_s\) normalizes \(\chi\); it is not an additional physical transport coefficient. The corresponding amplitudes are
    \begin{equation}
        \Theta = e^\tau,
        \qquad
        \Xi = e^\chi.
    \end{equation}
    The velocity becomes
    \begin{equation}
        \bm{u}
        =
        -2\mu_c \bm{\nabla}\tau
        +
        2\mu_s \bm{\nabla}^\perp\chi.
        \label{eq:u_tau_chi}
    \end{equation}
    Its divergence is
    \begin{equation}
        \bm{\nabla}\cdot \bm{u} = -2\mu_c \Delta \tau.
        \label{eq:div_u_tau_chi}
    \end{equation}
    Therefore \eqref{eq:continuity_s_general} becomes
    \begin{equation}
        \partial_t s + \bm{u}\cdot\bm{\nabla} s = 2\mu_c \Delta\tau. \label{eq:s_eq_tau}
    \end{equation}

\subsection{Compressive Schr\"odinger-type equation}
    Substituting \(\phi=-2\mu_c\tau\) into \eqref{eq:phi_eq} and using
    \begin{equation}
        \partial_t\Theta = \Theta\,\partial_t\tau,
        \qquad
        \Delta \Theta = \Theta\bigl(\Delta\tau + |\bm{\nabla}\tau|^2\bigr),
    \end{equation}
    we find
    \begin{equation}
        \partial_t\Theta = \mu_c \Delta \Theta + V_\Theta \Theta,
        \label{eq:Theta_heat}
    \end{equation}
    with scalar potential
    \begin{equation}
        V_\Theta
        =
        -2\mu_s\,\bm{\nabla}\tau\cdot\bm{\nabla}^\perp\chi
        +\frac{\mu_s^2}{\mu_c}|\bm{\nabla}\chi|^2
        +\frac{c_s^2}{2\mu_c}s
        +\frac{\Phi_F}{2\mu_c}.
        \label{eq:VTheta}
    \end{equation}

\subsection{Vortical Schr\"odinger-type equation}
    Substituting \(\psi=2\mu_s\chi\) into \eqref{eq:psi_eq}, and then writing \(\chi=\ln\Xi\), gives
    \begin{equation}
        \partial_t\Xi = \nu \Delta \Xi + V_\Xi \Xi,
        \label{eq:Xi_heat}
    \end{equation}
    where
    \begin{equation}
        V_\Xi
        =
        -\frac{\Psi_F}{2\mu_s}
        -\nu |\bm{\nabla}\chi|^2.
        \label{eq:VXi}
    \end{equation}

\subsection{The nonlocal forcing field}
    Since
    \begin{equation}
        \omega = -\Delta\psi = -2\mu_s \Delta\chi,
    \end{equation}
    and
    \begin{equation}
        \bm{u}^\perp
        =
        2\mu_c \bm{\nabla}^\perp \tau + 2\mu_s \bm{\nabla}\chi,
    \end{equation}
    we have
    \begin{equation}
        \bm{F} = \bm{u}^\perp \omega
        =
        -4\mu_s (\Delta\chi)\left(\mu_c \bm{\nabla}^\perp\tau + \mu_s \bm{\nabla}\chi\right).
        \label{eq:F_explicit}
    \end{equation}
    We determine the nonlocal potentials from
    \begin{equation}
        \Delta\Phi_F = \bm{\nabla}\cdot \bm{F},
        \qquad
        \Delta\Psi_F = - \bm{\nabla}^\perp\cdot \bm{F}.
        \label{eq:PhiPsi_def}
    \end{equation}

\subsection{Mixed density-compressive amplitude}
    We now seek an amplitude built from the density and compressive potential that has a second-order Schr\"odinger-type form. We choose the powers so that the explicit \(\Delta\tau\) and \(|\bm{\nabla}\tau|^2\) terms cancel from the reaction coefficient. Consider
    \begin{equation}
        \Psi_{\alpha,\beta} \equiv \rho^\alpha \Theta^\beta = e^{\alpha s + \beta \tau}.
    \end{equation}
    We write
    \begin{equation}
        q_{\alpha,\beta} \equiv \ln \Psi_{\alpha,\beta} = \alpha s + \beta\tau,
    \end{equation}
    then
    \begin{equation}
        (\partial_t+\bm{u}\cdot\bm{\nabla})\Psi_{\alpha,\beta}
        =
        \Psi_{\alpha,\beta}(\partial_t+\bm{u}\cdot\bm{\nabla})q_{\alpha,\beta}.
        \label{eq:Psiab_material}
    \end{equation}
    Using \eqref{eq:s_eq_tau} and
    \begin{equation}
        \partial_t\tau = \mu_c\Delta\tau + \mu_c|\bm{\nabla}\tau|^2 + V_\Theta,
    \label{eq:tau_t_from_Theta}
    \end{equation}
    which follows from \eqref{eq:Theta_heat}, we obtain
    \begin{align}
        (\partial_t+\bm{u}\cdot\bm{\nabla})q_{\alpha,\beta}
        &=
        \alpha(\partial_t+\bm{u}\cdot\bm{\nabla})s
        + \beta(\partial_t+\bm{u}\cdot\bm{\nabla})\tau
        \nonumber\\
        &=
        2\alpha\mu_c\Delta\tau
        + \beta\bigl(\partial_t\tau+\bm{u}\cdot\bm{\nabla}\tau\bigr)
        \nonumber\\
        &=
        \mu_c(2\alpha+\beta)\Delta\tau
        + \beta\mu_c|\bm{\nabla}\tau|^2
        \nonumber\\
        &\quad
        + \beta\bigl(V_\Theta+\bm{u}\cdot\bm{\nabla}\tau\bigr).
        \label{eq:qab_material}
    \end{align}
    On the other hand,
    \begin{equation}
        \bm{\nabla} q_{\alpha,\beta} = \alpha\bm{\nabla} s + \beta\bm{\nabla}\tau,
        \qquad
        \Delta q_{\alpha,\beta} = \alpha\Delta s + \beta\Delta\tau,
    \end{equation}
    so
    \begin{align}
        \Delta\Psi_{\alpha,\beta}
        &=
        \Psi_{\alpha,\beta}\bigl(\Delta q_{\alpha,\beta}+|\bm{\nabla} q_{\alpha,\beta}|^2\bigr)
        \nonumber\\
        &=
        \Psi_{\alpha,\beta}\Bigl(
        \alpha\Delta s + \beta\Delta\tau
        + \alpha^2 |\bm{\nabla} s|^2
        \nonumber\\
        &\quad
        + \beta^2 |\bm{\nabla}\tau|^2
        + 2\alpha\beta\,\bm{\nabla} s\cdot\bm{\nabla}\tau
        \Bigr).
        \label{eq:LapPsiab}
    \end{align}
    Substituting \eqref{eq:qab_material} and \eqref{eq:LapPsiab} into \eqref{eq:Psiab_material} gives
    \begin{equation}
        (\partial_t + \bm{u}\cdot\bm{\nabla})\Psi_{\alpha,\beta}
        =
        D\,\Delta\Psi_{\alpha,\beta}
        +
        U_{\alpha,\beta;D}\Psi_{\alpha,\beta},
        \label{eq:Psiab_advdiff_general}
    \end{equation}
    with
    \begin{align}
        U_{\alpha,\beta;D}
        &=
        \mu_c(2\alpha+\beta)\Delta\tau
        + \beta\mu_c|\bm{\nabla}\tau|^2
        \nonumber\\
        &\quad
        + \beta\bigl(V_\Theta+\bm{u}\cdot\bm{\nabla}\tau\bigr)
        - D\Bigl(
        \alpha\Delta s + \beta\Delta\tau
        + \alpha^2 |\bm{\nabla} s|^2
        \nonumber\\
        &\qquad
        + \beta^2 |\bm{\nabla}\tau|^2
        + 2\alpha\beta\,\bm{\nabla} s\cdot\bm{\nabla}\tau
        \Bigr).
        \label{eq:UabD_full}
    \end{align}
    Choosing the coefficients so that the explicit \(\Delta\tau\) and \(|\bm{\nabla}\tau|^2\) terms cancel yields
    \begin{equation}
        \beta = 1-2\alpha,
        \qquad
        D_\alpha = \frac{\mu_c}{1-2\alpha},
        \qquad
        \alpha \neq \frac12.
    \end{equation}
    Thus we define the mixed density-compressive amplitude
    \begin{equation}
        \Psi_\alpha \equiv \rho^\alpha \Theta^{1-2\alpha}.
    \end{equation}
    Equivalently,
    \begin{equation}
        \Psi_\alpha = e^{\alpha s + (1-2\alpha)\tau}.
    \end{equation}
    For a fixed \(\Theta\), any two choices \(\alpha,\alpha'\neq0,1/2\) are algebraically equivalent:
    \begin{equation}
        \Psi_{\alpha'}
        =
        \Psi_\alpha^{\alpha'/\alpha}
        \Theta^{1-\alpha'/\alpha}.
        \label{eq:Psi_alpha_reparam}
    \end{equation}
    Thus \(\alpha\) does not introduce a new physical field; it redistributes density and compressive-potential information between \(\Theta\) and \(\Psi_\alpha\). The restrictions \(\alpha\neq1/2\) and \(\alpha\neq0\) have different roles: \(\alpha\neq1/2\) keeps \(D_\alpha\) finite, while \(\alpha\neq0\) is needed to reconstruct \(s=\ln\rho\) from \(\Theta\) and \(\Psi_\alpha\).
    Its advection-diffusion equation is
    \begin{equation}
        (\partial_t+\bm{u}\cdot\bm{\nabla})\Psi_\alpha
        =
        D_\alpha \Delta\Psi_\alpha + U_\alpha \Psi_\alpha,
    \label{eq:Psi_alpha_advdiff}
    \end{equation}
    with
    \begin{align}
        U_\alpha
        &=
        -\frac{\mu_c\alpha}{1-2\alpha}\Delta s
        -\frac{\mu_c\alpha^2}{1-2\alpha}|\bm{\nabla} s|^2
        \nonumber\\
        &\quad
        -2\mu_c\alpha\,\bm{\nabla} s\cdot\bm{\nabla}\tau
        +(1-2\alpha)\bigl(V_\Theta+\bm{u}\cdot\bm{\nabla}\tau\bigr).
        \label{eq:Ualpha}
    \end{align}
    To absorb the advective derivative into a shifted Laplacian, rewrite \eqref{eq:Psi_alpha_advdiff} as
    \begin{equation}
        \partial_t\Psi_\alpha
        =
        D_\alpha\Delta\Psi_\alpha
        - \bm{u}\cdot\bm{\nabla}\Psi_\alpha
        + U_\alpha \Psi_\alpha.
        \label{eq:Psi_alpha_noncovariant}
    \end{equation}
    We define
    \begin{equation}
        \bm{\mathcal A}_\alpha \equiv -\frac{\bm{u}}{2D_\alpha}.
    \end{equation}
    Then \(2D_\alpha\bm{\mathcal A}_\alpha=-\bm{u}\), and for any scalar \(f\),
    \begin{align}
        D_\alpha(\bm{\nabla}+\bm{\mathcal A}_\alpha)^2 f
        &=
        D_\alpha\Delta f
        + 2D_\alpha\bm{\mathcal A}_\alpha\cdot\bm{\nabla} f
        \nonumber\\
        &\quad
        + D_\alpha\bigl(\bm{\nabla}\cdot\bm{\mathcal A}_\alpha
        + |\bm{\mathcal A}_\alpha|^2\bigr)f
        \nonumber\\
        &=
        D_\alpha\Delta f - \bm{u}\cdot\bm{\nabla} f
        \nonumber\\
        &\quad
        + D_\alpha\bigl(\bm{\nabla}\cdot\bm{\mathcal A}_\alpha
        + |\bm{\mathcal A}_\alpha|^2\bigr)f.
        \label{eq:covariant_identity}
    \end{align}
    Applying \eqref{eq:covariant_identity} to \eqref{eq:Psi_alpha_noncovariant} gives the vector-potential-coupled Schr\"odinger-type equation
    \begin{equation}
        \partial_t\Psi_\alpha
        =
        D_\alpha(\bm{\nabla}+\bm{\mathcal A}_\alpha)^2\Psi_\alpha
        + V_\alpha^{(v)}\Psi_\alpha,
        \label{eq:Psi_alpha_vector}
    \end{equation}
    where
    \begin{equation}
        V_\alpha^{(v)}
        =
        U_\alpha
        - D_\alpha\bigl(\bm{\nabla}\cdot\bm{\mathcal A}_\alpha + |\bm{\mathcal A}_\alpha|^2\bigr)
        =
        U_\alpha + \frac12 \bm{\nabla}\cdot\bm{u} - \frac{|\bm{u}|^2}{4D_\alpha}.
    \label{eq:Vv_alpha}
    \end{equation}
    Using \eqref{eq:u_tau_chi}, the vector potential is
    \begin{equation}
        \bm{\mathcal A}_\alpha
        =
        (1-2\alpha)\bm{\nabla}\tau
        -\frac{\mu_s(1-2\alpha)}{\mu_c}\,\bm{\nabla}^\perp\chi.
        \label{eq:CalAalpha_explicit}
    \end{equation}
    Equations \eqref{eq:Psi_alpha_vector}--\eqref{eq:CalAalpha_explicit} are the vector-potential-coupled Schr\"odinger-type equation for the transformed density variable \(\Psi_\alpha\). The operator is analogous to the magnetic Schr\"odinger operator, where a vector potential enters through the shifted derivative in the Laplacian \cite{AvronHerbstSimon1978_magnetic_schrodinger}. Here, however, \(\bm{\mathcal A}_\alpha\) is a real self-consistent flow potential rather than an imposed electromagnetic vector potential. It contains both compressive and vortical contributions and is not divergence-free in general. Consequently, \(D_\alpha(\bm{\nabla}+\bm{\mathcal A}_\alpha)^2\) is a real drift-diffusion operator and is generally not self-adjoint under periodic or decaying boundary conditions; its formal adjoint contains \((\bm{\nabla}-\bm{\mathcal A}_\alpha)^2\). This distinction is important when comparing the transformed density equation with Hermitian magnetic Schr\"odinger operators.

\subsection{Final coupled Eulerian system}
    We obtain a convenient nonlocal Eulerian closure by combining the two scalar amplitudes \(\Theta,\Xi\) with one selected member of the transformed density family, denoted here by
    \(\Psi_\alpha\), where \(\alpha\neq 0\) and \(\alpha\neq 1/2\),
    \begin{align}
        \partial_t\Theta &= \mu_c \Delta \Theta + V_\Theta \Theta,
        \label{eq:final_Theta}
        \\
        \partial_t\Xi &= \nu \Delta \Xi + V_\Xi \Xi,
        \label{eq:final_Xi}
        \\
        \partial_t\Psi_\alpha &= D_\alpha(\bm{\nabla}+\bm{\mathcal A}_\alpha)^2 \Psi_\alpha + V_\alpha^{(v)} \Psi_\alpha.
        \label{eq:final_Psi}
    \end{align}
    Here
    \begin{align}
        \Psi_\alpha &= \rho^\alpha \Theta^{1-2\alpha},
        &
        D_\alpha &= \frac{\mu_c}{1-2\alpha},
        \nonumber\\
        \bm{\mathcal A}_\alpha
        &=
        (1-2\alpha)\bm{\nabla}\tau
        -\frac{\mu_s(1-2\alpha)}{\mu_c}\,\bm{\nabla}^\perp\chi,
    \end{align}
    and the effective potentials are
    \begin{align}
        V_\Theta
        &=
        -2\mu_s\,\bm{\nabla}\tau\cdot\bm{\nabla}^\perp\chi
        +\frac{\mu_s^2}{\mu_c}|\bm{\nabla}\chi|^2
        +\frac{c_s^2}{2\mu_c}s
        +\frac{\Phi_F}{2\mu_c},
        \\
        V_\Xi
        &=
        -\frac{1}{2\mu_s}\Psi_F - \nu |\bm{\nabla}\chi|^2,
        \\
        V_\alpha^{(v)}
        &=
        -\frac{\mu_c\alpha}{1-2\alpha}\Delta s
        -\frac{\mu_c\alpha^2}{1-2\alpha}|\bm{\nabla} s|^2
        \nonumber\\
        &\quad
        -2\mu_c\alpha\,\bm{\nabla} s\cdot\bm{\nabla}\tau
        -\mu_c \Delta\tau
        \nonumber\\
        &\quad
        -3\mu_c(1-2\alpha)|\bm{\nabla}\tau|^2
        +2\mu_s(1-2\alpha)\,\bm{\nabla}\tau\cdot\bm{\nabla}^\perp\chi
        \nonumber\\
        &\quad
        +\frac{(1-2\alpha)c_s^2}{2\mu_c}s
        +\frac{1-2\alpha}{2\mu_c}\Phi_F .
    \end{align}
    We obtain the nonlocal quantities \(\Phi_F\) and \(\Psi_F\) from \eqref{eq:F_explicit}--\eqref{eq:PhiPsi_def}. We reconstruct the scalar fields as
    \begin{align}
        \tau &= \ln\Theta,
        &
        \chi &= \ln\Xi,
        \nonumber\\
        s=\ln\rho
        &=
        \frac{1}{\alpha}\Bigl(\ln \Psi_\alpha - (1-2\alpha)\tau\Bigr),
        &
        \alpha&\neq 0.
    \end{align}
    We reconstruct the velocity from
    \begin{equation}
        \bm{u}
        =
        -2\mu_c \bm{\nabla}\tau + 2\mu_s \bm{\nabla}^\perp\chi.
    \end{equation}
    In Appendix~\ref{app:obstruction}, we make explicit why pure density-only transforms do not close into a local Schr\"odinger-type equation for the transform classes considered here. In Appendix~\ref{app:other_transforms}, we show that broader density-compressive transform classes lead to the same restriction.

\section{Three-Dimensional Navier--Stokes}
\label{sec:three_dimensional_ns}
\subsection{Helmholtz decomposition in three dimensions}
    We now turn to a three-dimensional extension. We replace the scalar streamfunction used in two dimensions by a divergence-free vector streamfunction, and the scalar vorticity by the vorticity vector. We write the velocity in Helmholtz form
    \begin{equation}
        \bm{u} = \bm{\nabla} \phi + \bm{\nabla}\times \bm{\psi},
        \qquad
        \bm{\nabla}\cdot\bm{\psi}=0,
    \end{equation}
    so that
    \begin{equation}
        \bm{\nabla}\cdot\bm{u} = \Delta\phi,
        \qquad
        \bm{\omega} \equiv \bm{\nabla}\times\bm{u} = -\Delta\bm{\psi},
    \end{equation}
    where the second identity uses the Coulomb gauge. The condition \(\bm{\nabla}\cdot\bm{\psi}=0\) is a gauge choice for the vector streamfunction rather than an additional physical constraint, since \(\bm{\psi}\mapsto\bm{\psi}+\bm{\nabla}q\) leaves \(\bm{\nabla}\times\bm{\psi}\) unchanged. The continuity equation again becomes
    \begin{equation}
        \partial_t s + \bm{u}\cdot\bm{\nabla} s = -\Delta\phi.
        \label{eq:continuity_s_3d}
    \end{equation}

\subsection{Gradient projection and compressive amplitude}
    In three dimensions,
    \begin{equation}
        \bm{u}\cdot\bm{\nabla}\bm{u}
        =
        \bm{\nabla}\!\left(\frac{|\bm{u}|^2}{2}\right) - \bm{u}\times\bm{\omega}.
    \end{equation}
    We define
    \begin{equation}
        \bm{G} \equiv -\bm{u}\times\bm{\omega}.
    \end{equation}
    We decompose \(\bm{G}\) into gradient and solenoidal parts,
    \begin{equation}
        \bm{G} = \bm{\nabla} \Phi_G + \bm{\nabla}\times \bm{Y}_G,
        \qquad
        \bm{\nabla}\cdot\bm{Y}_G = 0.
    \end{equation}
    Projecting the momentum equation onto the gradient sector gives
    \begin{equation}
        \partial_t\phi + \frac{|\bm{u}|^2}{2} + c_s^2 s + \Phi_G = \mu_c \Delta\phi,
        \qquad
        \mu_c \equiv \zeta+2\nu.
        \label{eq:phi_eq_3d}
    \end{equation}
    We set
    \begin{equation}
        \phi = -2\mu_c\tau,
        \qquad
        \Theta = e^\tau,
    \end{equation}
    and denote the divergence-free part by
    \begin{equation}
        \bm{u}_s \equiv \bm{\nabla}\times\bm{\psi},
        \qquad
        \bm{\nabla}\cdot\bm{u}_s = 0.
    \end{equation}
    Then
    \begin{equation}
        \bm{u} = -2\mu_c\bm{\nabla}\tau + \bm{u}_s,
        \label{eq:u_tau_us_3d}
    \end{equation}
    and \eqref{eq:continuity_s_3d} becomes
    \begin{equation}
        \partial_t s + \bm{u}\cdot\bm{\nabla} s = 2\mu_c\Delta\tau.
        \label{eq:s_eq_tau_3d}
    \end{equation}
    The same logarithmic calculation used in the two-dimensional compressive sector yields
    \begin{equation}
        \partial_t\Theta = \mu_c\Delta\Theta + V_{\Theta}^{(3d)}\Theta,
        \label{eq:Theta_3d}
    \end{equation}
    with
    \begin{equation}
        V_{\Theta}^{(3d)}
        =
        -\bm{\nabla}\tau\cdot\bm{u}_s
        +\frac{|\bm{u}_s|^2}{4\mu_c}
        +\frac{c_s^2}{2\mu_c}s
        +\frac{\Phi_G}{2\mu_c}.
        \label{eq:VTheta_3d}
    \end{equation}

\subsection{Transformed density amplitude in three dimensions}
    The transformed density variable has the same density-compressive form as in two dimensions:
    \begin{align}
        \Psi_\alpha \equiv \rho^\alpha \Theta^{1-2\alpha}
        &=
        e^{\alpha s + (1-2\alpha)\tau},
        \nonumber\\
        D_\alpha
        &=
        \frac{\mu_c}{1-2\alpha},
        \qquad
        \alpha\neq \frac12.
    \end{align}
    The same cancellation mechanism as in two dimensions gives
    \begin{align}
        (\partial_t+\bm{u}\cdot\bm{\nabla})\Psi_\alpha
        &=
        D_\alpha\Delta\Psi_\alpha + U_\alpha^{(3d)}\Psi_\alpha,
        \label{eq:Psi_alpha_advdiff_3d}
    \end{align}
    where
    \begin{align}
        U_\alpha^{(3d)}
        &=
        -\frac{\mu_c\alpha}{1-2\alpha}\Delta s
        -\frac{\mu_c\alpha^2}{1-2\alpha}|\bm{\nabla} s|^2
        \nonumber\\
        &\quad
        -2\mu_c\alpha\,\bm{\nabla} s\cdot\bm{\nabla}\tau
        +(1-2\alpha)\bigl(V_{\Theta}^{(3d)}+\bm{u}\cdot\bm{\nabla}\tau\bigr).
        \label{eq:Ualpha_3d}
    \end{align}
    Equivalently,
    \begin{align}
        \partial_t\Psi_\alpha
        &=
        D_\alpha\Delta\Psi_\alpha
        - \bm{u}\cdot\bm{\nabla}\Psi_\alpha
        + U_\alpha^{(3d)}\Psi_\alpha.
    \end{align}
    We introduce the effective vector potential
    \begin{equation}
        \bm{\mathcal A}_\alpha^{(3d)}
        \equiv
        -\frac{\bm{u}}{2D_\alpha}
        =
        (1-2\alpha)\bm{\nabla}\tau - \frac{1-2\alpha}{2\mu_c}\,\bm{u}_s.
        \label{eq:Aalpha_3d}
    \end{equation}
    Then the transformed density variable satisfies the three-dimensional vector-potential-coupled Schr\"odinger-type equation
    \begin{equation}
        \partial_t\Psi_\alpha
        =
        D_\alpha(\bm{\nabla}+\bm{\mathcal A}_\alpha^{(3d)})^2\Psi_\alpha
        +V_\alpha^{(v,3d)}\Psi_\alpha,
        \label{eq:Psi_alpha_vector_3d}
    \end{equation}
    with
    \begin{equation}
        V_\alpha^{(v,3d)}
        =
        U_\alpha^{(3d)}
        +\frac12\bm{\nabla}\cdot\bm{u}
        -\frac{|\bm{u}|^2}{4D_\alpha}.
        \label{eq:Vv_3d_compact}
    \end{equation}
    Using \eqref{eq:u_tau_us_3d}, this becomes
    \begin{align}
        V_\alpha^{(v,3d)}
        &=
        -\frac{\mu_c\alpha}{1-2\alpha}\Delta s
        -\frac{\mu_c\alpha^2}{1-2\alpha}|\bm{\nabla} s|^2
        \nonumber\\
        &\quad
        -2\mu_c\alpha\,\bm{\nabla} s\cdot\bm{\nabla}\tau
        -\mu_c\Delta\tau
        \nonumber\\
        &\quad
        -3\mu_c(1-2\alpha)|\bm{\nabla}\tau|^2
        +(1-2\alpha)\,\bm{\nabla}\tau\cdot\bm{u}_s
        \nonumber\\
        &\quad
        +\frac{(1-2\alpha)c_s^2}{2\mu_c}s
        +\frac{1-2\alpha}{2\mu_c}\Phi_G.
        \label{eq:Vv_3d_explicit}
    \end{align}
    Thus the transformed density equation has the same vector-potential-coupled structure in three dimensions as in two dimensions.

\subsection{Solenoidal sector and vector-streamfunction transform}
    For incompressible three-dimensional flow, Ohkitani showed that the solenoidal vector-potential equation can be rewritten component-wise in Cole--Hopf form \cite{Ohkitani2017,VanonOhkitani2018}. We write the corresponding velocity potential as the vector streamfunction \(\bm{\psi}\), reserving \(\bm{A}\) for the magnetic vector potential in Appendix~\ref{app:MHD}. This construction builds on the earlier vector-potential evolution equations for incompressible Euler and NS flow \cite{Ohkitani2015_PRE}. We treat the compressible case by the same sequence of steps after projecting the momentum equation onto its solenoidal sector. We start from the compressible momentum equation written in the form
    \begin{align}
        \partial_t \bm{u}
        &+ \bm{\nabla}\!\left(\frac{|\bm{u}|^2}{2}\right)
        + c_s^2\bm{\nabla} s
        + \bm{G}
        \nonumber\\
        &= (\mu_c-\nu)\bm{\nabla}(\bm{\nabla}\cdot\bm{u})
        + \nu\Delta\bm{u},
        \nonumber\\
	        \bm{G}&=-\bm{u}\times\bm{\omega}.
	        \label{eq:momentum_3d_split}
	    \end{align}
    We denote the Leray projector onto divergence-free vector fields by \(\mathbb{P}\).
    The projector is defined by \cite{Leray1934}
    \begin{equation}
        \mathbb{P}\bm{F}
        =
        \bm{F}-\bm{\nabla}\Delta^{-1}(\bm{\nabla}\cdot\bm{F}),
        \label{eq:Leray_projector}
    \end{equation}
    so \(\bm{\nabla}\cdot(\mathbb{P}\bm{F})=0\) and \(\bm{F}\) splits into a gradient part plus its solenoidal projection. In Fourier variables,
    \begin{equation}
        \widehat{\mathbb{P}\bm{F}}(\bm{k})
        =
        \left(I-\frac{\bm{k}\otimes\bm{k}}{|\bm{k}|^2}\right)\widehat{\bm{F}}(\bm{k}).
    \end{equation}
    Applying \(\mathbb{P}\) annihilates all gradient terms. Since \(\mathbb{P}\bm{u}=\bm{u}_s\), \(\mathbb{P}\bm{\nabla}(\bm{\nabla}\cdot\bm{u})=0\), and \(\mathbb{P}\Delta\bm{u}=\Delta\bm{u}_s\), we obtain
    \begin{equation}
        \partial_t \bm{u}_s + \mathbb{P}\bm{G} = \nu\Delta\bm{u}_s.
        \label{eq:us_proj_3d}
    \end{equation}
    Using the Helmholtz decomposition \(\bm{G}=\bm{\nabla}\Phi_G+\bm{\nabla}\times\bm{Y}_G\), the projected nonlinear term is \(\mathbb{P}\bm{G}=\bm{\nabla}\times\bm{Y}_G\). Therefore
    \begin{equation}
        \partial_t \bm{u}_s + \bm{\nabla}\times\bm{Y}_G = \nu\Delta\bm{u}_s.
        \label{eq:us_eq_3d}
    \end{equation}
    Since \(\bm{u}_s=\bm{\nabla}\times\bm{\psi}\), \eqref{eq:us_eq_3d} determines the vector streamfunction only up to a gradient gauge. We now show that one can choose this gauge, compatibly with the periodic or decaying boundary conventions used to define the inverse Laplacian, so that
    \begin{equation}
        \partial_t \bm{\psi} + \bm{Y}_G = \nu\Delta\bm{\psi},
        \qquad
        \bm{\nabla}\cdot\bm{\psi}=0.
        \label{eq:psi_vec_eq_3d_compressible}
    \end{equation}
    Substituting \(\bm{u}_s=\bm{\nabla}\times\bm{\psi}\) into \eqref{eq:us_eq_3d} gives
    \begin{equation}
        \bm{\nabla}\times\Bigl(\partial_t\bm{\psi}+\bm{Y}_G-\nu\Delta\bm{\psi}\Bigr)=0.
    \end{equation}
    Thus the residual is a pure gradient: there exists a scalar field \(\lambda\) such that
    \begin{equation}
    \partial_t\bm{\psi}+\bm{Y}_G-\nu\Delta\bm{\psi}=\bm{\nabla}\lambda.
    \end{equation}
    We use the gauge transform \(\bm{\psi}\mapsto \bm{\psi}+\bm{\nabla} q\), with
    \begin{equation}
        \partial_t q-\nu\Delta q=-\lambda,
    \end{equation}
    which preserves \(\bm{u}_s=\bm{\nabla}\times\bm{\psi}\) and removes the pure-gradient residual. In this gauge \(\lambda=0\), yielding \eqref{eq:psi_vec_eq_3d_compressible}. This evolution preserves the Coulomb gauge: because \(\bm{\nabla}\cdot\bm{Y}_G=0\), taking the divergence gives \(\partial_t(\bm{\nabla}\cdot\bm{\psi})=\nu\Delta(\bm{\nabla}\cdot\bm{\psi})\). Thus \(\bm{\nabla}\cdot\bm{\psi}=0\) remains true if imposed initially under the periodic or decaying boundary conditions considered here. In the incompressible limit, this reduces to Ohkitani's
    vector-streamfunction equation. The forcing is determined by the full compressible field
    \begin{align}
        \bm{u} &= -2\mu_c\bm{\nabla}\tau + \bm{\nabla}\times\bm{\psi},
        \\
        \bm{\omega} &= \bm{\nabla}\times\bm{u} = -\Delta\bm{\psi},
        \\
        \bm{G} &= -\bm{u}\times\bm{\omega}.
    \end{align}
    Substituting \(\bm{u}=-2\mu_c\bm{\nabla}\tau+\bm{u}_s\) gives the explicit split
    \begin{equation}
        \bm{G}
        =
        -\bm{u}_s\times\bm{\omega}
        +2\mu_c\,\bm{\nabla}\tau\times\bm{\omega}.
        \label{eq:G_split_3d}
    \end{equation}
    The first term is the incompressible Ohkitani contribution, while the second is the genuine compressible correction. Accordingly,
    \begin{equation}
        \mathbb{P}\bm{G}
        =
        \mathbb{P}(-\bm{u}_s\times\bm{\omega})
        +2\mu_c\,\mathbb{P}(\bm{\nabla}\tau\times\bm{\omega}).
        \label{eq:PG_split_3d}
    \end{equation}
    Choosing \(\bm{\nabla}\cdot\bm{Y}_G=0\), we write
    \begin{equation}
        \bm{Y}_G=\bm{Y}_{s}+\bm{Y}_{c},
        \label{eq:YG_split_3d}
    \end{equation}
    where
    \begin{equation}
        \bm{\nabla}\times\bm{Y}_{s}
        =
        \mathbb{P}(-\bm{u}_s\times\bm{\omega}),
        \quad
        \bm{\nabla}\times\bm{Y}_{c}
        =
        2\mu_c\,\mathbb{P}(\bm{\nabla}\tau\times\bm{\omega}).
        \label{eq:YG_curls_3d}
    \end{equation}
    Equivalently,
    \begin{align}
        \bm{Y}_{s}
        &=
        -\Delta^{-1}\bm{\nabla}\times(-\bm{u}_s\times\bm{\omega}),
        \nonumber\\
        \bm{Y}_{c}
        &=
        -2\mu_c\,\Delta^{-1}\bm{\nabla}\times(\bm{\nabla}\tau\times\bm{\omega}),
        \label{eq:YG_explicit_3d}
    \end{align}
    since \(\bm{\nabla}\times(\mathbb{P}\bm{F})=\bm{\nabla}\times\bm{F}\) and \(\bm{\nabla}\times(\bm{\nabla}\times\bm{Y})=-\Delta\bm{Y}\) under the Coulomb gauge. Thus the compressible effect on the solenoidal sector is encoded in the additional nonlocal term \(\bm{Y}_{c}\). Applying the component-wise Cole--Hopf transform
    \begin{equation}
        \psi_j = k\ln \vartheta_j,
        \qquad
        j=1,2,3,
        \label{eq:psij_transform_3d_compressible}
    \end{equation}
    where \(k\) is a transform scale, gives
    \begin{equation}
        \partial_t \vartheta_j
        =
        \nu\Delta\vartheta_j + V_{\vartheta,j}^{(3d)}\vartheta_j,
        \qquad
        \text{(no summation)},
        \label{eq:vartheta_j_3d}
    \end{equation}
    with
    \begin{equation}
        V_{\vartheta,j}^{(3d)}
        =
        -\frac{(Y_G)_j}{k}
        -\nu\frac{|\bm{\nabla}\vartheta_j|^2}{\vartheta_j^2}.
        \label{eq:Vvartheta_j_3d}
    \end{equation}
    This follows component-wise from
    \begin{equation}
        k\frac{\partial_t\vartheta_j}{\vartheta_j} + (Y_G)_j
        =
        \nu k\left(
        \frac{\Delta\vartheta_j}{\vartheta_j}
        - \frac{|\bm{\nabla}\vartheta_j|^2}{\vartheta_j^2}
        \right),
    \end{equation}
    after substituting \(\psi_j=k\ln\vartheta_j\) into \eqref{eq:psi_vec_eq_3d_compressible} and using \(\Delta\ln\vartheta_j=\Delta\vartheta_j/\vartheta_j-|\bm{\nabla}\vartheta_j|^2/\vartheta_j^2\). The Coulomb-gauge constraint becomes
    \begin{equation}
        \sum_{j=1}^3 \frac{1}{\vartheta_j}\partial_j \vartheta_j = 0.
        \label{eq:vartheta_constraint_3d}
    \end{equation}
    We summarize the coupled transformed system below.

\subsection{Coupled transformed system in three dimensions}
    Collecting the preceding results, we write the compressible analogue of Ohkitani's three-dimensional formulation as
    \begin{align}
        \partial_t\Theta &= \mu_c\Delta\Theta + V_{\Theta}^{(3d)}\Theta,
        \label{eq:final_Theta_3d}
        \\
        \partial_t\Psi_\alpha
        &=
        D_\alpha(\bm{\nabla}+\bm{\mathcal A}_\alpha^{(3d)})^2\Psi_\alpha
        +V_\alpha^{(v,3d)}\Psi_\alpha,
        \label{eq:final_Psi_3d}
        \\
        \partial_t \vartheta_j
        &=
        \nu\Delta\vartheta_j + V_{\vartheta,j}^{(3d)}\vartheta_j,
        \qquad
        j=1,2,3.
        \label{eq:final_vartheta_3d}
    \end{align}
    There is no summation over \(j\) in \eqref{eq:final_vartheta_3d}. Here
    \begin{equation}
        V_{\Theta}^{(3d)}
        =
        -\bm{\nabla}\tau\cdot\bm{u}_s
        +\frac{|\bm{u}_s|^2}{4\mu_c}
        +\frac{c_s^2}{2\mu_c}s
        +\frac{\Phi_G}{2\mu_c},
    \end{equation}
    \begin{equation}
        \bm{\mathcal A}_\alpha^{(3d)}
        =
        (1-2\alpha)\bm{\nabla}\tau - \frac{1-2\alpha}{2\mu_c}\,\bm{u}_s,
    \end{equation}
    \begin{align}
        V_\alpha^{(v,3d)}
        &=
        -\frac{\mu_c\alpha}{1-2\alpha}\Delta s
        -\frac{\mu_c\alpha^2}{1-2\alpha}|\bm{\nabla} s|^2
        \nonumber\\
        &\quad
        -2\mu_c\alpha\,\bm{\nabla} s\cdot\bm{\nabla}\tau
        -\mu_c\Delta\tau
        \nonumber\\
        &\quad
        -3\mu_c(1-2\alpha)|\bm{\nabla}\tau|^2
        +(1-2\alpha)\,\bm{\nabla}\tau\cdot\bm{u}_s
        \nonumber\\
        &\quad
        +\frac{(1-2\alpha)c_s^2}{2\mu_c}s
        +\frac{1-2\alpha}{2\mu_c}\Phi_G,
    \end{align}
    and
    \begin{equation}
        V_{\vartheta,j}^{(3d)}
        =
        -\frac{(Y_G)_j}{k}
        -\nu\frac{|\bm{\nabla}\vartheta_j|^2}{\vartheta_j^2}.
    \end{equation}
    We reconstruct the physical variables from
    \begin{equation}
        \tau = \ln\Theta,
        \qquad
        s = \frac{1}{\alpha}\Bigl(\ln\Psi_\alpha - (1-2\alpha)\tau\Bigr),
        \qquad
        \rho = e^s,
    \end{equation}
    \begin{equation}
        \psi_j = k\ln\vartheta_j,
        \qquad
        \bm{u}_s = \bm{\nabla}\times\bm{\psi},
        \qquad
        \bm{u} = -2\mu_c\bm{\nabla}\tau + \bm{u}_s,
    \end{equation}
    with the Coulomb-gauge constraint
    \begin{equation}
        \sum_{j=1}^3 \frac{1}{\vartheta_j}\partial_j\vartheta_j = 0.
    \end{equation}
    These reconstruction formulas require \(\alpha\neq 0\) and \(\alpha\neq 1/2\). Together, they give the coupled compressive-solenoidal-density transformed system for the present Eulerian compressible generalization.
    
    \begin{figure*}[tbp]
        \includegraphics[width=\textwidth]{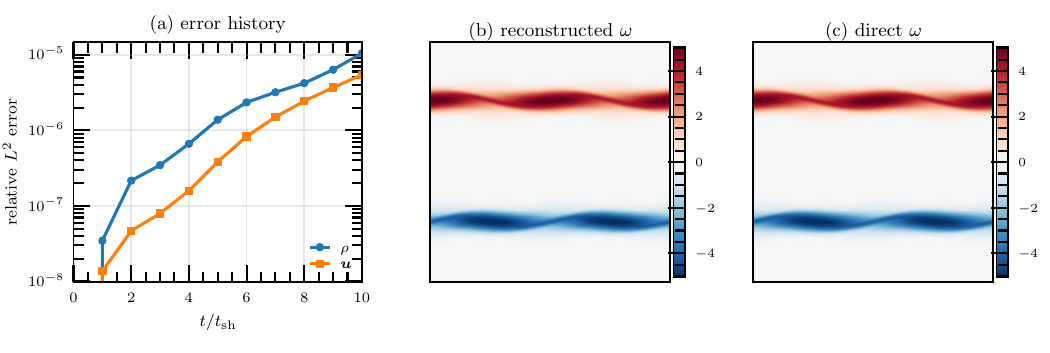}
        \caption{Kelvin--Helmholtz validation of the transformed system at \(128^2\). Panel (a) shows the relative \(L^2\) errors in the reconstructed density \(\rho\) and velocity \(\bm{u}\) compared with the direct compressible Navier--Stokes simulation. The errors grow smoothly through the shear-layer evolution but remain at \(\mathcal{O}(10^{-5})\) or below at \(t/t_{\rm sh}=10.00\). Panels (b) and (c) compare the final reconstructed and direct vorticity fields, showing that the two rolled shear layers and their peak vorticity amplitudes are recovered without
        visible phase drift.}
        \label{fig:khi_validation}
    \end{figure*}

    \begin{figure*}[tbp]
        \includegraphics[width=\textwidth]{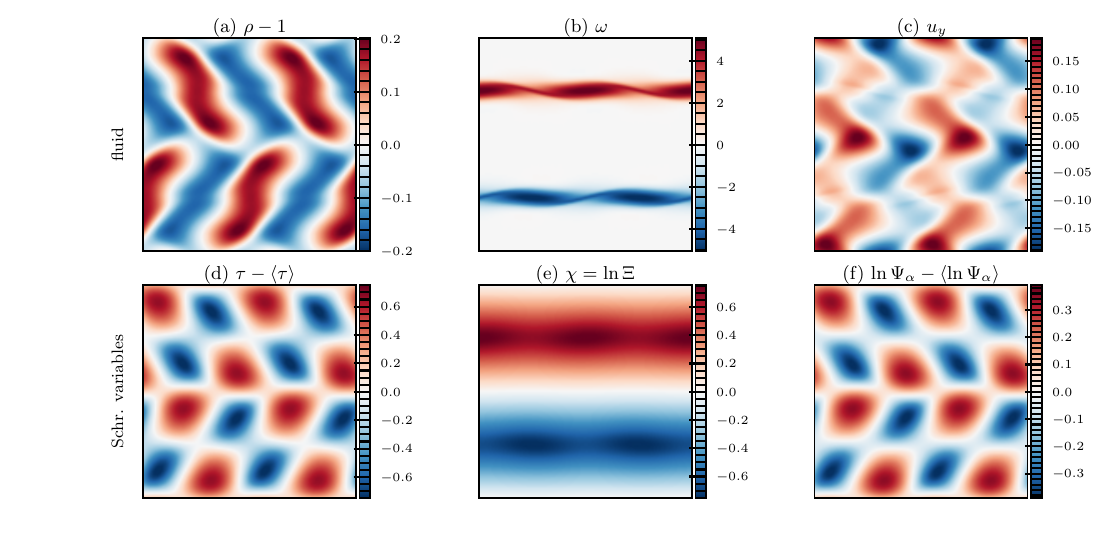}
        \caption{Final Kelvin--Helmholtz state in fluid and transformed variables at \(t/t_{\rm sh}=10.00\). The top row shows the reconstructed physical fields: The density perturbation \(\rho-1\), the vorticity \(\omega\), and the transverse velocity \(u_y\). The bottom row shows the logarithmic transformed variables \(\tau=\ln\Theta\), \(\chi=\ln\Xi\), and \(\ln\Psi_\alpha\), with spatial means subtracted from \(\tau\) and \(\ln\Psi_\alpha\). The transformed fields separate different pieces of the same flow: \(\chi\) tracks the two vorticity-bearing shear layers, while \(\tau\) and \(\ln\Psi_\alpha\) carry the larger-scale compressive and density-coupled structure.}
        \label{fig:khi_schrodinger_fluid}
    \end{figure*}

    \begin{figure*}[tbp]
        \includegraphics[width=\textwidth]{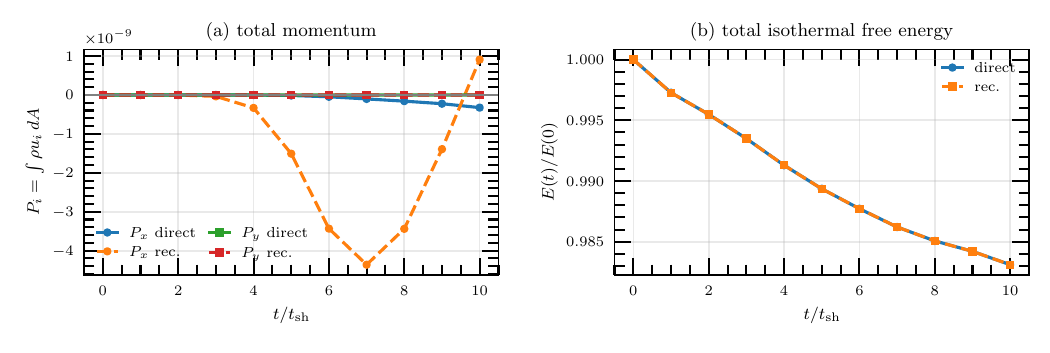}
        \caption{Integral diagnostics for the same \(128^2\) Kelvin--Helmholtz comparison. Panel (a) shows the domain-integrated momenta \(P_i=\int\rho u_i\,dA\) for the direct and reconstructed states. Both components remain close to their initial values, with the small residuals shown on a \(10^{-9}\) scale. Panel (b) shows the normalized isothermal free energy \(E=\int [\rho|\bm{u}|^2/2+c_s^2(\rho\ln\rho-\rho+1)]\,dA\). The energy decays because of viscosity, and the direct and reconstructed curves track the same decay.}
        \label{fig:khi_integral_diagnostics}
    \end{figure*}

\section{Numerical validation}
\label{sec:numerical_validation}
    We validate the two-dimensional transformed system \eqref{eq:final_Theta}--\eqref{eq:final_Psi} by evolving it alongside a direct simulation of the compressible isothermal equations \eqref{eq:continuity_rho}--\eqref{eq:momentum_u} on a doubly periodic square domain \((x,y)\in [0,2\pi]^2\). The calculations use \textsc{dedalus}, an open-source parallel framework for solving partial differential equations with spectral methods \cite{Burns2020_Dedalus}. Here \textsc{dedalus} provides the Fourier discretization for the direct simulation and the periodic Poisson solver used in the transformed equations. The transformed state consists of \(\Theta\), \(\Xi\), and \(\Psi_\alpha\), with \(\Phi_F\) and \(\Psi_F\) obtained at each time by solving the zero-mean Poisson problems in \eqref{eq:PhiPsi_def},
    \begin{equation}
        \Delta\Phi_F = \bm{\nabla}\cdot\bm{F},
        \qquad
        \Delta\Psi_F = -\bm{\nabla}^\perp\cdot\bm{F}.
    \end{equation}
    The physical variables are reconstructed point-wise from
    \begin{align}
        \tau &= \ln\Theta,
        &
        \chi &= \ln\Xi,
        \nonumber\\
        s &=
        \frac{1}{\alpha}\Bigl(\ln\Psi_\alpha-(1-2\alpha)\tau\Bigr),
        &
        \rho&=e^s,
        \label{eq:validation_scalar_reconstruction}
    \end{align}
    and
    \begin{equation}
        \bm{u}
        =
        -2\mu_c\bm{\nabla}\tau + 2\mu_s\bm{\nabla}^\perp\chi.
        \label{eq:validation_velocity_reconstruction}
    \end{equation}
    The logarithmic reconstruction requires positivity of the transformed amplitudes; loss of positivity would signal either numerical failure or departure from the regime in which the amplitude variables are well conditioned.

    The test problem is a smooth Kelvin--Helmholtz unstable shear layer, a standard instability (the KHI) problem that has also been used to test potential-based formulations of incompressible flow \cite{Ohkitani2015_PRE}. It is a good test because it has analytical growth rate and gets into nonlinear regime where the numerical methods are stringently tested. The initial density is uniform, \(\rho(x,y,0)=1\), and the velocity is initialized as
    \begin{align}
        u_x(x,y,0)
        &=
        U_0\bigl[P(y)-\langle P\rangle\bigr],
        \\
        u_y(x,y,0)
        &=
        \epsilon\sin(k_x x)E(y),
        \label{eq:khi_initial_velocity}
    \end{align}
    where
    \begin{align}
        E(y)
        &=
        e^{-[(y-y_1)/\sigma_p]^2}
        +e^{-[(y-y_2)/\sigma_p]^2},
        \\
        P(y)&=
        \tanh\frac{y-y_1}{\sigma_s}-\tanh\frac{y-y_2}{\sigma_s}-1 .
    \end{align}
    The parameters are \(y_1=\pi/2\), \(y_2=3\pi/2\), \(k_x=2\), \(U_0=1\), and \(\epsilon=0.5\), with shear width \(\sigma_s=0.03(2\pi)\) and perturbation width \(\sigma_p=0.06(2\pi)\). We report times in units of the KHI shear time
    \begin{equation}
        t_{\rm sh}\equiv \frac{\sigma_s}{U_0}=0.1885,
    \end{equation}
    which is the inverse peak shear of an isolated \(\tanh\) layer with the present normalization. The direct and transformed systems are initialized from the same transformed state, so the reconstructed fields \eqref{eq:validation_scalar_reconstruction}--\eqref{eq:validation_velocity_reconstruction} and direct physical variables agree to roundoff at \(t/t_{\rm sh}=0\).

    The calculation used \(128^2\) Fourier collocation points, \(c_s=1\), \(\nu=5\times10^{-4}\), \(\zeta=5\times10^{-2}\), \(\alpha=1/4\), \(\mu_s=1\), and a fixed timestep \(\Delta t=5\times10^{-4}=2.65\times10^{-3}t_{\rm sh}\). The final time is \(t/t_{\rm sh}=10.00\). The direct reference simulation was advanced with the \textsc{dedalus} RK443 timestepper. The transformed system \eqref{eq:final_Theta}--\eqref{eq:final_Psi} used an explicit RK4 update, with nonlinear products evaluated on the \(3/2\)-dealiased grid and \textsc{dedalus} Poisson-solver calls for the projected potentials \eqref{eq:PhiPsi_def}. Each transformed right-hand-side evaluation requires two elliptic solver calls, one for \(\Phi_F\) and one for \(\Psi_F\). The RK4 update therefore uses eight periodic Poisson-solver calls per transformed timestep, in addition to the two one-time Poisson-solver calls used to initialize \(\tau\) and \(\chi\) from the physical velocity field. Thus the errors below measure the implemented side-by-side validation, including timestepper, projection, and elliptic-solver errors, rather than a formal convergence rate.

    \begin{table}[t]
    \caption{Relative-error diagnostics for the \(128^2\) Kelvin--Helmholtz comparison. Times are reported in KHI shear units, \(t_{\rm sh}=\sigma_s/U_0\). The density and logarithmic density errors \(\epsilon_\rho\) and \(\epsilon_s\) are relative \(L^2\) norms of \(\rho_{\rm rec}-\rho_{\rm dir}\) and \(s_{\rm rec}-s_{\rm dir}\), respectively. The velocity error \(\epsilon_u\) is the relative \(L^2\) norm of \(|\bm{u}_{\rm rec}-\bm{u}_{\rm dir}|\).}
    \label{tab:khi_validation}
    \begin{ruledtabular}
    \begin{tabular}{c c c c}
        \(t/t_{\rm sh}\) & \(\epsilon_\rho\) & \(\epsilon_s\) & \(\epsilon_u\)\\
        \hline
        0.00 & 0 & 0 & 0\\
        2.00 & \(2.16\times10^{-7}\) & \(2.41\times10^{-6}\) &
        \(4.66\times10^{-8}\)\\
        4.00 & \(6.62\times10^{-7}\) & \(6.22\times10^{-6}\) &
        \(1.58\times10^{-7}\)\\
        6.00 & \(2.34\times10^{-6}\) & \(1.94\times10^{-5}\) &
        \(8.28\times10^{-7}\)\\
        8.00 & \(4.21\times10^{-6}\) & \(3.20\times10^{-5}\) &
        \(2.43\times10^{-6}\)\\
        10.00 & \(1.05\times10^{-5}\) & \(9.49\times10^{-5}\) &
        \(5.42\times10^{-6}\)\\
    \end{tabular}
    \end{ruledtabular}
    \end{table}

    Table~\ref{tab:khi_validation} gives the relative-error diagnostics. The errors increase monotonically as the shear layers roll up, but remain small through the final time. At \(t/t_{\rm sh}=10.00\), the density and velocity relative \(L^2\) errors are \(1.05\times10^{-5}\) and \(5.42\times10^{-6}\), respectively. The logarithmic density error \(\epsilon_s\) is larger than \(\epsilon_\rho\), as expected because \(s=\ln\rho\) weights fractional density differences; even so, it remains below \(10^{-4}\). These numbers are not a convergence study, but they show that the implemented transformed right-hand side \eqref{eq:final_Theta}--\eqref{eq:final_Psi}, projection operations \eqref{eq:PhiPsi_def}, and reconstruction formulas \eqref{eq:validation_scalar_reconstruction}--\eqref{eq:validation_velocity_reconstruction} remain consistent with the direct simulation over several turnover times.

    Fig.~\ref{fig:khi_validation} gives a complementary visual check. The error history in panel (a) shows no sudden loss of agreement or positivity, and panels (b) and (c) show that the reconstructed and direct vorticity fields have the same phase, layer thickness, and extrema at the final time. This comparison is important because small global \(L^2\) errors alone could miss a coherent phase shift of the rolled shear layers.

    The amplitude variables remain positive throughout the run, as required by the logarithmic reconstruction \eqref{eq:validation_scalar_reconstruction}--\eqref{eq:validation_velocity_reconstruction}. At the final time \(\min\rho_{\rm rec}=0.829781\), \(\min\rho_{\rm dir}=0.829784\), \(\min\Theta=1.53\times10^{3}\), \(\min\Xi=0.4687\), and \(\min\Psi_\alpha=38.43\). Fig.~\ref{fig:khi_schrodinger_fluid} shows the same final KHI state in both the ordinary fluid variables and the logarithmic transformed variables. Here \(\tau\) is proportional to the compressive velocity potential, \(\phi=-2\mu_c\tau\), so its spatial structure reflects the irrotational part of the velocity field. By contrast, \(\ln\Psi_\alpha=\alpha s+(1-2\alpha)\tau\) combines the logarithmic density \(s=\ln\rho\) with the compressive potential. The difference between the \(\tau\) and \(\ln\Psi_\alpha\) panels therefore shows how density fluctuations modulate the compressive sector during the shear-layer roll-up. The comparison illustrates what the transformation does geometrically: \(\chi=\ln\Xi\) carries the two vorticity-bearing shear layers, while \(\tau=\ln\Theta\) and \(\ln\Psi_\alpha\) encode broader compressive and density-coupled structure. Thus the transformed variables are not merely auxiliary numerical fields; they separate the vortical, compressive, and density information that is mixed together in \((\rho,\bm{u})\).

    In Fig.~\ref{fig:khi_integral_diagnostics} we check the integrated quantities. The total momentum components are
    \begin{equation}
        P_i(t)=\int \rho u_i\,dA,
        \qquad i=x,y.
        \label{eq:momentum_validation}
    \end{equation}
    For periodic conservative compressible flow these are the natural momentum diagnostics, and the components in \eqref{eq:momentum_validation} remain close to their initial zero values in both evolutions: over the saved snapshots, the maximum absolute drift is \(3.3\times10^{-10}\) for the direct \(P_x\), \(4.4\times10^{-9}\) for the reconstructed \(P_x\), and below \(6.0\times10^{-15}\) and \(2.6\times10^{-15}\) for the direct and reconstructed \(P_y\), respectively.

    The energy diagnostic requires a small distinction. Isothermal compressible flow does not conserve kinetic energy on its own, because pressure work exchanges kinetic energy with compressive free energy. The density contribution below is the standard relative pressure potential for barotropic compressible flow \cite{Feireisl2004_viscous_compressible}: it vanishes at the reference state \(\rho=1\) and, for small density perturbations, is quadratic in \(\rho-1\). For periodic inviscid isothermal flow the corresponding conserved functional is
    \begin{equation}
        E(t)=\int
        \left[
            \frac12\rho|\bm{u}|^2
            +c_s^2\bigl(\rho\ln\rho-\rho+1\bigr)
        \right]\,dA,
        \label{eq:isothermal_free_energy_validation}
    \end{equation}
    where the second term is the isothermal internal/free-energy density, normalized to vanish at \(\rho=1\). In the present viscous calculation the functional \eqref{eq:isothermal_free_energy_validation} is expected to decay rather than remain constant. It decreases to \(E(t)/E(0)=0.9831\) by \(t/t_{\rm sh}=10.00\). The direct and reconstructed normalized energy histories differ by at most \(2.9\times10^{-10}\), providing an additional global check that the transformed simulation follows the same dissipative evolution as the direct simulation.

    \begin{table*}[tbp]
    \caption{Summary of the compressible NS transformation to imaginary-time Schr\"odinger-type amplitude equations. The table separates the variables used in the closed two-dimensional system \eqref{eq:final_Theta}--\eqref{eq:final_Psi} from their three-dimensional analogues in \eqref{eq:final_Theta_3d}--\eqref{eq:final_vartheta_3d}, and lists where the self-consistent potentials entering each transformed equation are defined.}
    \label{tab:operator_structure}
    \renewcommand{\arraystretch}{1.18}
    \begin{ruledtabular}
    \begin{tabular}{p{0.14\textwidth}@{\hspace{0.018\textwidth}}p{0.21\textwidth}@{\hspace{0.018\textwidth}}p{0.25\textwidth}@{\hspace{0.018\textwidth}}p{0.27\textwidth}}
        \ltab{Variable} & \ltab{Physical content} & \ltab{Leading transformed equation} & \ltab{Explicit potentials and nonlocal input}\\
        \hline
        \ltab{\begin{tabular}[t]{@{}l@{}}2D/3D\\\(\Theta=e^\tau\)\end{tabular}} & \ltab{compressive potential \(\phi=-2\mu_c\tau\), hence \(\bm{\nabla}\cdot\bm{u}=-2\mu_c\Delta\tau\) in 2D \eqref{eq:div_u_tau_chi}; the same relation follows in 3D from \eqref{eq:u_tau_us_3d}} & \ltab{2D: \(\partial_t\Theta=\mu_c\Delta\Theta+V_\Theta\Theta\) in \eqref{eq:final_Theta}\par\vspace{0.45ex}3D: \(\partial_t\Theta=\mu_c\Delta\Theta+V_\Theta^{(3d)}\Theta\) in \eqref{eq:final_Theta_3d}} & \ltab{\(V_\Theta\) in \eqref{eq:VTheta} contains pressure \(c_s^2s\), vortical coupling through \(\chi\), and \(\Phi_F\); \(V_\Theta^{(3d)}\) in \eqref{eq:VTheta_3d} replaces the vortical scalar coupling by \(\bm{u}_s\) and uses \(\Phi_G\). The nonlocal scalar potentials satisfy \eqref{eq:PhiPsi_def}}\\
        \ltab{\begin{tabular}[t]{@{}l@{}}2D scalar\\streamfunction\\\(\Xi=e^\chi\)\end{tabular}} & \ltab{scalar streamfunction \(\psi=2\mu_s\chi\), hence \(\omega=-2\mu_s\Delta\chi\)} & \ltab{\(\partial_t\Xi=\nu\Delta\Xi+V_\Xi\Xi\) in \eqref{eq:final_Xi}} & \ltab{\(V_\Xi\) in \eqref{eq:VXi} contains the projected vortical forcing \(\Psi_F\) and logarithmic-gradient term \(|\bm{\nabla}\chi|^2\); \(\Psi_F\) is obtained with \(\Phi_F\) from \eqref{eq:F_explicit}--\eqref{eq:PhiPsi_def}}\\
        \ltab{\begin{tabular}[t]{@{}l@{}}3D vector\\streamfunction\\\(\vartheta_j=e^{\psi_j/k}\)\end{tabular}} & \ltab{vector streamfunction \(\bm{u}_s=\bm{\nabla}\times\bm{\psi}\), with \(\psi_j=k\ln\vartheta_j\); in Coulomb gauge, \(\bm{\omega}=-\Delta\bm{\psi}\)} & \ltab{component-wise equation \(\partial_t\vartheta_j=\nu\Delta\vartheta_j+V_{\vartheta,j}^{(3d)}\vartheta_j\) in \eqref{eq:final_vartheta_3d}} & \ltab{\(V_{\vartheta,j}^{(3d)}\) is defined in \eqref{eq:Vvartheta_j_3d}. It contains the projected solenoidal forcing \((Y_G)_j\), where \(\bm{Y}_G=\bm{Y}_s+\bm{Y}_c\) in \eqref{eq:YG_split_3d}; \(\bm{Y}_s\) is the incompressible-type contribution and \(\bm{Y}_c\) is the compressible correction driven by \(\bm{\nabla}\tau\), as in \eqref{eq:YG_curls_3d}--\eqref{eq:YG_explicit_3d}}\\
        \ltab{\begin{tabular}[t]{@{}l@{}}2D/3D\\\(\Psi_\alpha=\rho^\alpha\Theta^{1-2\alpha}\)\end{tabular}} & \ltab{density-compressive amplitude, with \(s=\ln\rho=\alpha^{-1}[\ln\Psi_\alpha-(1-2\alpha)\tau]\)} & \ltab{2D: \(\partial_t\Psi_\alpha=D_\alpha(\bm{\nabla}+\bm{\mathcal A}_\alpha)^2\Psi_\alpha+V_\alpha^{(v)}\Psi_\alpha\) in \eqref{eq:final_Psi}\par\vspace{0.45ex}3D: \(\partial_t\Psi_\alpha=D_\alpha(\bm{\nabla}+\bm{\mathcal A}_\alpha^{(3d)})^2\Psi_\alpha+V_\alpha^{(v,3d)}\Psi_\alpha\) in \eqref{eq:final_Psi_3d}} & \ltab{2D uses \(\bm{\mathcal A}_\alpha\) in \eqref{eq:CalAalpha_explicit} and \(V_\alpha^{(v)}\) in \eqref{eq:Vv_alpha}; 3D uses \(\bm{\mathcal A}_\alpha^{(3d)}\) in \eqref{eq:Aalpha_3d} and \(V_\alpha^{(v,3d)}\) in \eqref{eq:Vv_3d_explicit}. These potentials contain density gradients, pressure, compressive gradients, solenoidal coupling, and the nonlocal scalars \(\Phi_F\) or \(\Phi_G\)}\\
    \end{tabular}
    \end{ruledtabular}
    \end{table*}

\section{Discussion and Conclusion}
\label{sec:discussion}
\label{sec:conclusion}
    The main result of this paper is an exact Eulerian Cole--Hopf-type operator reformulation of the 2D and 3D isothermal compressible Navier--Stokes (NS) equations and its extension to MHD. To our knowledge, this is the first exact Cole--Hopf-type Schr\"odinger-variable reformulation of compressible NS flow. In Sec.~\ref{sec:two_dimensional_ns}, the system \eqref{eq:final_Theta}--\eqref{eq:final_Psi} gives the closed transformed dynamics in terms of three amplitudes:
    \begin{enumerate}
            \item a scalar compressive amplitude \(\Theta\) satisfying the heat or imaginary-time Schr\"odinger-type equation \eqref{eq:final_Theta},
            \item a scalar vortical amplitude \(\Xi\) satisfying the corresponding equation \eqref{eq:final_Xi},
            \item a scalar density-carrying amplitude \(\Psi_\alpha\) satisfying the vector-potential-coupled Schr\"odinger-type equation \eqref{eq:final_Psi}, with effective vector potential \(\bm{\mathcal A}_\alpha\) given by \eqref{eq:CalAalpha_explicit}.
    \end{enumerate}
    The density sector does not close as a pure density amplitude; instead, the mixed amplitude \(\Psi_\alpha=\rho^\alpha\Theta^{1-2\alpha}\), with \(\alpha\neq 0,\frac12\), cancels the compressive Laplacian terms and closes in the vector-potential-coupled form. The resulting system is exact, but nonlocal: the Helmholtz decomposition and the projected nonlinear forcing require Poisson-solver calls, with the forcing potentials determined by \eqref{eq:PhiPsi_def}. Table~\ref{tab:operator_structure} summarizes the bookkeeping used in this discussion. Its first two columns identify the physical content of each amplitude, its third column lists the corresponding transformed evolution equation, and its final column locates the self-consistent potentials that carry the nonlinear and nonlocal couplings. The \(\Theta\), \(\Xi\), and \(\Psi_\alpha\) rows give the closed 2D amplitude system, while the \(\Theta\), \(\Psi_\alpha\), and \(\vartheta_j\) rows show how the same operator structure extends to the 3D compressive, density-carrying, and solenoidal sectors.

    The Kelvin--Helmholtz calculation in Sec.~\ref{sec:numerical_validation} validates this transformed 2D system against a direct compressible NS simulation for the test considered here. At \(t/t_{\rm sh}=10.00\), the reconstructed \(\rho\) and \(\bm{u}\) differ from the direct fields by relative \(L^2\) errors \(1.05\times 10^{-5}\) and \(5.42\times10^{-6}\), respectively, and the reconstructed vorticity field remains phase-aligned with the direct simulation (Fig.~\ref{fig:khi_validation}). The transformed fields in Fig.~\ref{fig:khi_schrodinger_fluid} show how the logarithmic variables separate vortical, compressive, and density-coupled structure. The amplitudes satisfy \(\Theta,\Xi,\Psi_\alpha>0\) throughout the run, the direct and reconstructed momenta \(P_i(t)\) agree to small absolute drift, and the normalized isothermal free energies \(E(t)/E(0)\) differ by at most \(2.9\times10^{-10}\) (Fig.~\ref{fig:khi_integral_diagnostics}). These diagnostics check the implemented transformed right-hand side, nonlocal projections, and reconstruction formulas; they should not be read as a convergence study or as evidence that the transformed variables automatically improve numerical stability.

    The construction has three immediate limitations. First, the logarithmic reconstruction requires \(\Theta,\Xi,\Psi_\alpha>0\), so positivity must be monitored in any numerical use of the transformed variables. Second, \(\rho\) cannot be represented by a single pure density amplitude satisfying a closed second-order Schr\"odinger-type equation within the local point-transform classes considered here. We make this restriction explicit in Appendix~\ref{app:obstruction} and derive related transform classes in Appendix~\ref{app:other_transforms}. Third, we assume an isothermal equation of state, \(p=c_s^2\rho\), which closes the mass--momentum system by prescribing the pressure directly from \(\rho\) rather than by evolving an independent energy or entropy equation; extending the formulation to fully adiabatic compressible flow would require transforming or evolving this additional thermodynamic degree of freedom. Evolving the energy or entropy equation would be necessary for any relativistic reformulation, where the equation of state must keep the sound speed causally bounded \(c_s\le c\) by the speed of light, $c$.

    In Sec.~\ref{sec:three_dimensional_ns}, we show that the 3D transformed density equation keeps the same vector-potential-coupled structure, \eqref{eq:final_Psi_3d}, once \(\bm{u}\) is split into compressive and solenoidal parts. As summarized in Table~\ref{tab:operator_structure}, the 3D vector-streamfunction amplitudes \(\vartheta_j\) satisfy component-wise heat or imaginary-time Schr\"odinger-type equations with multiplicative potentials \(V_{\vartheta,j}^{(3d)}\). The solenoidal sector follows the vector-streamfunction strategy used for incompressible NS flow by Ohkitani and by Vanon and Ohkitani \cite{Ohkitani2017,VanonOhkitani2018}, with an additional nonlocal term \(\bm{Y}_{c}\) driven by \(\bm{\nabla}\tau\) in \eqref{eq:YG_explicit_3d}. In Appendix~\ref{app:MHD}, we show that the same projected-potential viewpoint also extends to compressible isothermal MHD, where the Lorentz force modifies the solenoidal velocity forcing and the induction equation gives a companion magnetic-potential transform. These 3D and MHD results therefore give novel formal exact changes of variables for compressible systems, while leaving open whether the new variables yield bounds, regularity criteria, stable numerical schemes, or computational advantages.

    The transformed operators may also suggest reduced descriptions of the flow. Each snapshot defines self-consistent operators \(\mathcal L_\Theta=\mu_c\Delta+V_\Theta\), \(\mathcal L_\Xi=\nu\Delta+V_\Xi\), and \(\mathcal L_{\Psi_\alpha}=D_\alpha(\bm{\nabla}+\bm{\mathcal A}_\alpha)^2+V_\alpha^{(v)}\), associated with \eqref{eq:final_Theta}--\eqref{eq:final_Psi}, with \(V_\Theta\), \(V_\Xi\), \(\bm{\mathcal A}_\alpha\), and \(V_\alpha^{(v)}\) evaluated from the instantaneous reconstructed solution. Their instantaneous eigenfunctions could be tested as adaptive bases in the spirit of modal decompositions and projection-based reduced-order models \cite{Holmes2012_turbulence,Benner2015_model_reduction,Taira2017_modal}. This is only a possible use of the formulation: the operators are nonlinear functionals of the solution, \(\mathcal L_{\Psi_\alpha}\) need not be self-adjoint, and it remains to determine whether such bases are more compact or more informative than Fourier, proper orthogonal decomposition, or other standard modal representations.

    The same explicit potentials also suggest a way to build controlled truncated models. The final column of Table~\ref{tab:operator_structure} separates the scalar multiplicative potentials, the vector-potential-coupled density terms, and the nonlocal forcing potentials. In 2D, the nonlinear couplings enter through \(V_\Theta\), \(V_\Xi\), and \(V_\alpha^{(v)}\), defined in \eqref{eq:VTheta}, \eqref{eq:VXi}, and \eqref{eq:Vv_alpha}, through the vector potential \(\bm{\mathcal A}_\alpha\) in \eqref{eq:CalAalpha_explicit}, and through \(\Phi_F,\Psi_F\) from \eqref{eq:F_explicit}--\eqref{eq:PhiPsi_def}. In the original fluid variables, these elliptic forcing potentials are the compressive and vortical Helmholtz projections of \(\bm{F}=\bm{u}^\perp\omega\), the vorticity-coupled part of nonlinear advection after the kinetic-energy gradient has been separated. For example, setting \(\Phi_F=\Psi_F=0\) while retaining the local pressure, density-gradient, and logarithmic-gradient terms would define a nonlocal-forcing-free amplitude model that tests how this advective vorticity force feeds back on the transformed amplitudes. Retaining or suppressing selected pieces of these potentials would break exact equivalence to compressible NS, but it provides a systematic way to isolate the nonlinear compressive, vortical, density-coupled, and nonlocal projection effects.

    For quantum-computing applications, this reformulation should not be read as a quantum speedup claim. Rather, it rewrites the compressible isothermal NS equations as a coupled set of amplitude equations whose heat- and Schr\"odinger-type linear parts may be able to benefit from quantum algorithms for operator evolution, in the same limited sense explored for transformed Burgers and heat-equation problems \cite{Uchida2024_Burgers_quantum,Linden2022_heat_equation}. The closest quantum-fluid comparison is the HSE of Meng and Yang \cite{MengYang2023_hydrodynamic_schrodinger}: their two-component wave function gives a unitary Schr\"odinger-type evolution and motivates a prediction-correction quantum algorithm, whereas our logarithmic amplitudes give imaginary-time heat or drift-diffusion equations with self-consistent potentials. The tradeoff is also different. Their incompressible Schr\"odinger flow is designed to resemble important structures of viscous turbulence; our construction remains exactly equivalent to isothermal compressible NS, but only by retaining nonlocal Poisson projections, nonlinear potentials, and positivity constraints. More broadly, the present variables sit alongside quantum and quantum-inspired approaches based on lattice-Boltzmann or flow-specific algorithms \cite{Budinski2022_quantum_nse,Succi2023_quantum_fluids,Wang2025_quantum_lbm}, hydrodynamic or Madelung-type Schr\"odinger reformulations \cite{Madelung1927,Zylberman2022_madelung_hydro,MengYang2023_hydrodynamic_schrodinger}, and variational or tensor-network representations of nonlinear PDEs \cite{MoczSzasz2021_quantum_cosmo,YeLoureiro2022_mps_vlasov,Tennie2025_quantum_nonlinear_turbulence}. The remaining obstacles are explicit: nonlinear self-consistent potentials, elliptic projections, positivity of logarithmic amplitudes, state preparation, and observable extraction. Recent quantum-hardware studies suggest that statistical quantities may be more accessible readouts than full pointwise fields \cite{Goldack2026_velocity_stats,Uchida2024_Burgers_quantum}, but determining whether these amplitude systems lead to useful analysis, reduced models, improved algorithms, or practical quantum implementations remains an open problem.

\section*{Data and code availability}
    The \textsc{dedalus} scripts used to generate the Kelvin--Helmholtz validation simulations and manuscript figures are available in Ref.~\cite{Beattie2026_dedalus_CH}. The validation data reported in Sec.~\ref{sec:numerical_validation} and Figs.~\ref{fig:khi_validation}--\ref{fig:khi_integral_diagnostics} can be regenerated from these scripts; no additional experimental data were used.

\section*{Acknowledgments}
    We thank Elias R. Most for the many enlightening discussions. We thank the two anonymous reviewers for comments that helped improve the clarity and presentation of the study. The authors used OpenAI Codex \cite{OpenAI2025_Codex} for assistance with \textsc{dedalus} code organization and manuscript editing. We acknowledge support from the Natural Sciences \& Engineering Research Council of Canada (NSERC, funding reference number 568580), the Canadian Space Agency (23JWGO2A01), the Simons Foundation (MP-SCMPS-00001470), NSF PHY-2309135 to the Kavli Institute for Theoretical Physics (KITP), and NSF Award 2206756. We acknowledge compute allocations rrg-ripperda and rrg-essick from the Digital Research Alliance of Canada, which were used to run and analyze the simulations, and high-performance computing resources provided by the Leibniz Rechenzentrum and the Gauss Center for Supercomputing (grants pn76gi, pr73fi, and pn76ga).

\appendix

\section{Equivalence of Leray projection and Green-function operators}
\label{app:green_operator}
    We relate the Leray-projector formulation used in Sec.~\ref{sec:three_dimensional_ns} to Ohkitani's Green-function operator \(T\). They are not different reductions; rather, \(T\) is the explicit integral realization of the projected nonlinear term after inverting curl and Laplacian.
    
    Let \(\bm{F}\) be a sufficiently regular vector field on \(\mathbb{R}^3\). The Leray projection \cite{Leray1934} is
    \begin{equation}
        \mathbb{P}\bm{F}
        =
        \bm{F}-\bm{\nabla}\Delta^{-1}(\bm{\nabla}\cdot\bm{F}),
    \end{equation}
    so \(\bm{\nabla}\cdot(\mathbb{P}\bm{F})=0\). Choose \(\bm{Y}_F\) in the Coulomb gauge, \(\bm{\nabla}\cdot\bm{Y}_F=0\), and require
    \begin{equation}
        \bm{\nabla}\times\bm{Y}_F = \mathbb{P}\bm{F},
        \label{eq:curlYF_app}
    \end{equation}
    Applying curl once more gives
    \begin{align}
        \bm{\nabla}\times(\bm{\nabla}\times\bm{Y}_F)
        &=
        \bm{\nabla}(\bm{\nabla}\cdot\bm{Y}_F)-\Delta\bm{Y}_F
        \nonumber\\
        &=
        -\Delta\bm{Y}_F
        =
        \bm{\nabla}\times(\mathbb{P}\bm{F})
        =
        \bm{\nabla}\times\bm{F},
    \end{align}
    because curl annihilates gradients. Therefore
    \begin{equation}
        \bm{Y}_F = -\Delta^{-1}\bm{\nabla}\times\bm{F}.
        \label{eq:YF_green_app}
    \end{equation}
    Applying this identity to the vortical sector, the projected equation
    \begin{equation}
        \partial_t \bm{u}_s + \mathbb{P}\bm{G} = \nu\Delta\bm{u}_s,
        \qquad
        \bm{u}_s=\bm{\nabla}\times\bm{\psi},
    \end{equation}
    is equivalent, up to gauge, to
    \begin{equation}
        \partial_t\bm{\psi} + \bm{Y}_G = \nu\Delta\bm{\psi},
        \qquad
        \bm{Y}_G = -\Delta^{-1}\bm{\nabla}\times\bm{G}.
        \label{eq:A_BG_equiv_app}
    \end{equation}
    Hence
    \begin{equation}
        \partial_t\bm{\psi}
        =
        \nu\Delta\bm{\psi} + \Delta^{-1}\bm{\nabla}\times\bm{G}.
        \label{eq:A_TG_app}
    \end{equation}
    We identify the Green-function operator associated with \(\bm{G}\) as
    \begin{equation}
        \bm{T}_G[\bm{G}] \equiv \Delta^{-1}\bm{\nabla}\times\bm{G}.
        \label{eq:TG_def_app}
    \end{equation}
    
    In the incompressible case, \(\bm{u}=\bm{u}_s=\bm{\nabla}\times\bm{\psi}\) and \(\bm{G}=-\bm{u}\times\bm{\omega}\). Then \eqref{eq:A_TG_app} becomes
    \begin{align}
        \partial_t\bm{\psi}
        &=
        \bm{T}_{s}[\bm{\nabla}\bm{\psi}] + \nu\Delta\bm{\psi},
        \nonumber\\
        \bm{T}_{s}[\bm{\nabla}\bm{\psi}]
        &\equiv
        \Delta^{-1}\bm{\nabla}\times(-\bm{u}\times\bm{\omega}),
        \label{eq:Tinc_app}
    \end{align}
    which is Ohkitani's vector-potential equation written in the present \(\bm{\psi}\) notation. Ohkitani's operator \(\bm{T}[\bm{\nabla}\bm{\psi}]\) is the explicit principal-value kernel representation of \(\bm{T}_{s}[\bm{\nabla}\bm{\psi}]\).

    In the compressible case,
    \begin{equation}
        \bm{u}=-2\mu_c\bm{\nabla}\tau+\bm{u}_s,
        \qquad
        \bm{G}
        =
        -\bm{u}_s\times\bm{\omega}
        +2\mu_c\,\bm{\nabla}\tau\times\bm{\omega},
    \end{equation}
    so the Green-function operator splits as
    \begin{equation}
        \bm{T}_G
        =
        \bm{T}_{s}[\bm{\nabla}\bm{\psi}]
        +\bm{T}_{c}[\tau,\bm{\psi}],
    \end{equation}
    with
    \begin{equation}
        \bm{T}_{c}[\tau,\bm{\psi}]
        \equiv
        2\mu_c\,\Delta^{-1}\bm{\nabla}\times(\bm{\nabla}\tau\times\bm{\omega}).
        \label{eq:Tcomp_app}
    \end{equation}
    Thus the Leray-projector formulation in Sec.~\ref{sec:three_dimensional_ns} and Ohkitani's \(T\)-operator formulation are equivalent descriptions of the same solenoidal dynamics. Our compressible generalization leaves the operator structure unchanged and adds the Green-function term driven by \(\bm{\nabla}\tau\).

\section{Compressible MHD extension}
\label{app:MHD}
    Vanon and Ohkitani give the incompressible MHD analogue of the vector-potential Cole--Hopf formulation in their appendix \cite{VanonOhkitani2018}. Here we derive the corresponding compressible isothermal system. We use units in which the magnetic permeability is one and write
    \begin{align}
        \partial_t\rho+\bm{\nabla}\cdot(\rho\bm{u}) &= 0,
        \label{eq:mhd_continuity_app}
        \\
        \partial_t\bm{u}+\bm{u}\cdot\bm{\nabla}\bm{u}
        &=
        -c_s^2\bm{\nabla}s
        +\nu\Delta\bm{u}
        \nonumber\\
        &\quad
        +(\zeta+\nu)\bm{\nabla}(\bm{\nabla}\cdot\bm{u}) \\
        &\qquad
        +e^{-s}\bm{J}\times\bm{B},
        \label{eq:mhd_momentum_app}
        \\
        \partial_t\bm{B}
        &=
        \bm{\nabla}\times(\bm{u}\times\bm{B})
        +\eta\Delta\bm{B},
        \label{eq:mhd_induction_app}
        \\
        \bm{\nabla}\cdot\bm{B} &= 0,
    \end{align}
    where \(s=\ln\rho\), \(\bm{B}\) is the magnetic field, \(\bm{J}=\bm{\nabla}\times\bm{B}\), and \(\eta\) is the magnetic diffusivity. The induction equation is equivalently
    \begin{equation}
        \partial_t\bm{B}
        +\bm{u}\cdot\bm{\nabla}\bm{B}
        =
        \bm{B}\cdot\bm{\nabla}\bm{u}
        -\bm{B}\,\bm{\nabla}\cdot\bm{u}
        +\eta\Delta\bm{B},
        \label{eq:mhd_induction_advective_app}
    \end{equation}
    using \(\bm{\nabla}\cdot\bm{B}=0\).

    As in Sec.~\ref{sec:three_dimensional_ns}, let
    \begin{align}
        \bm{u}&=-2\mu_c\bm{\nabla}\tau+\bm{u}_s,
        &
        \bm{u}_s&=\bm{\nabla}\times\bm{\psi},
        \nonumber\\
        \bm{\nabla}\cdot\bm{\psi}&=0,
        &
        \mu_c&=\zeta+2\nu,
        \label{eq:mhd_u_split_app}
    \end{align}
    and introduce a magnetic vector potential
    \begin{equation}
        \bm{B}=\bm{\nabla}\times\bm{A},
        \qquad
        \bm{\nabla}\cdot\bm{A}=0.
        \label{eq:mhd_A_def_app}
    \end{equation}
    Then \(\bm{\omega}=\bm{\nabla}\times\bm{u}=-\Delta\bm{\psi}\) and \(\bm{J}=-\Delta\bm{A}\). We define the Lorentz acceleration and the total non-gradient forcing in the velocity equation by
    \begin{equation}
        \bm{L}\equiv e^{-s}\bm{J}\times\bm{B},
        \qquad
        \bm{K}\equiv -\bm{u}\times\bm{\omega}-\bm{L}.
        \label{eq:mhd_K_def_app}
    \end{equation}
    Using \(\bm{u}\cdot\bm{\nabla}\bm{u}=\bm{\nabla}(|\bm{u}|^2/2)-\bm{u}\times\bm{\omega}\), the momentum equation becomes
    \begin{equation}
        \partial_t\bm{u}
        +\bm{\nabla}\!\left(\frac{|\bm{u}|^2}{2}+c_s^2s\right)
        +\bm{K}
        =
        (\mu_c-\nu)\bm{\nabla}(\bm{\nabla}\cdot\bm{u})
        +\nu\Delta\bm{u}.
        \label{eq:mhd_momentum_K_app}
    \end{equation}
    We decompose \(\bm{K}\) into gradient and solenoidal parts,
    \begin{equation}
        \bm{K}=\bm{\nabla}\Phi_K+\mathbb{P}\bm{K},
        \qquad
        \mathbb{P}\bm{K}=\bm{\nabla}\times\bm{Y}_K,
        \qquad
        \bm{\nabla}\cdot\bm{Y}_K=0.
        \label{eq:mhd_K_decomp_app}
    \end{equation}
    Equivalently,
    \begin{equation}
        \bm{Y}_K=-\Delta^{-1}\bm{\nabla}\times\bm{K}.
        \label{eq:mhd_YK_app}
    \end{equation}
    The gradient projection gives
    \begin{align}
        \partial_t\phi+\frac{|\bm{u}|^2}{2}+c_s^2s+\Phi_K&=\mu_c\Delta\phi,
        \nonumber\\
        \phi&=-2\mu_c\tau,
        &
        \Theta&=e^\tau,
    \end{align}
    hence
    \begin{equation}
        \partial_t\Theta=\mu_c\Delta\Theta+V_{\Theta}^{(\mathrm{MHD})}\Theta,
        \label{eq:mhd_Theta_app}
    \end{equation}
    with
    \begin{equation}
        V_{\Theta}^{(\mathrm{MHD})}
        =
        -\bm{\nabla}\tau\cdot\bm{u}_s
        +\frac{|\bm{u}_s|^2}{4\mu_c}
        +\frac{c_s^2}{2\mu_c}s
        +\frac{\Phi_K}{2\mu_c}.
        \label{eq:mhd_VTheta_app}
    \end{equation}
    Thus the density-carrying amplitude
    \begin{equation}
        \Psi_\alpha=\rho^\alpha\Theta^{1-2\alpha},
        \qquad
        D_\alpha=\frac{\mu_c}{1-2\alpha},
        \qquad
        \alpha\neq\frac12,
        \label{eq:mhd_Psi_def_app}
    \end{equation}
    obeys the same vector-potential-coupled form as in the hydrodynamic case,
    \begin{equation}
        \partial_t\Psi_\alpha
        =
        D_\alpha\bigl(\bm{\nabla}+\bm{\mathcal A}_\alpha^{(\mathrm{MHD})}\bigr)^2\Psi_\alpha
        +V_\alpha^{(v,\mathrm{MHD})}\Psi_\alpha,
        \label{eq:mhd_Psi_vector_app}
    \end{equation}
    where
    \begin{equation}
        \bm{\mathcal A}_\alpha^{(\mathrm{MHD})}
        =
        (1-2\alpha)\bm{\nabla}\tau
        -\frac{1-2\alpha}{2\mu_c}\bm{u}_s,
        \label{eq:mhd_Aalpha_app}
    \end{equation}
    and \(V_\alpha^{(v,\mathrm{MHD})}\) follows from \eqref{eq:Vv_3d_explicit} after replacing \(\Phi_G\) with \(\Phi_K\).

    The solenoidal velocity projection gives
    \begin{equation}
        \partial_t\bm{u}_s+\mathbb{P}\bm{K}=\nu\Delta\bm{u}_s,
    \end{equation}
    or, after the same gauge choice used in Sec.~\ref{sec:three_dimensional_ns},
    \begin{equation}
        \partial_t\bm{\psi}+\bm{Y}_K=\nu\Delta\bm{\psi}.
        \label{eq:mhd_psi_eq_app}
    \end{equation}
    With the component-wise logarithmic transform
    \begin{equation}
        \psi_j=k\ln\vartheta_j,
        \qquad
        j=1,2,3,
    \end{equation}
    we obtain
    \begin{align}
        \partial_t\vartheta_j
        &=
        \nu\Delta\vartheta_j
        +V_{\vartheta,j}^{(\mathrm{MHD})}\vartheta_j,
        \nonumber\\
        V_{\vartheta,j}^{(\mathrm{MHD})}
        &=
        -\frac{(Y_K)_j}{k}
        -\nu\frac{|\bm{\nabla}\vartheta_j|^2}{\vartheta_j^2},
        \label{eq:mhd_vartheta_app}
    \end{align}
    with no summation over \(j\), and the Coulomb constraint becomes
    \begin{equation}
        \sum_{j=1}^3\frac{1}{\vartheta_j}\partial_j\vartheta_j=0.
        \label{eq:mhd_vartheta_constraint_app}
    \end{equation}

    We now transform the induction equation. We define
    \begin{equation}
        \bm{W}\equiv\bm{u}\times\bm{B}.
    \end{equation}
    Since \(\partial_t\bm{B}=\bm{\nabla}\times\bm{W}+\eta\Delta\bm{B}\) and \(\bm{B}=\bm{\nabla}\times\bm{A}\), the magnetic vector potential satisfies
    \begin{equation}
        \partial_t\bm{A}
        =
        \mathbb{P}\bm{W}
        +\eta\Delta\bm{A}
        =
        \bm{W}+\bm{S}_W+\eta\Delta\bm{A},
        \label{eq:mhd_A_eq_app}
    \end{equation}
    where
    \begin{equation}
        \bm{S}_W
        \equiv
        -\bm{\nabla}\Delta^{-1}(\bm{\nabla}\cdot\bm{W})
        \label{eq:mhd_SW_app}
    \end{equation}
    is the gradient term that enforces the Coulomb gauge. We set
    \begin{equation}
        A_j=\ell\ln\mathcal M_j,
        \qquad
        j=1,2,3.
    \end{equation}
    Then
    \begin{align}
        \partial_t\mathcal M_j
        &=
        \eta\Delta\mathcal M_j
        +H_j^{(\mathrm{MHD})}\mathcal M_j,
        \nonumber\\
        H_j^{(\mathrm{MHD})}
        &=
        \frac{W_j+(S_W)_j}{\ell}
        -\eta\frac{|\bm{\nabla}\mathcal M_j|^2}{\mathcal M_j^2},
        \label{eq:mhd_M_app}
    \end{align}
    again with no summation over \(j\), and
    \begin{equation}
        \sum_{j=1}^3\frac{1}{\mathcal M_j}\partial_j\mathcal M_j=0.
        \label{eq:mhd_M_constraint_app}
    \end{equation}

    We therefore obtain the transformed compressible MHD system as the coupled set \eqref{eq:mhd_Theta_app}, \eqref{eq:mhd_Psi_vector_app}, \eqref{eq:mhd_vartheta_app}, and \eqref{eq:mhd_M_app}, together with the nonlocal definitions of \(\Phi_K\), \(\bm{Y}_K\), and \(\bm{S}_W\). We reconstruct the physical variables from
    \begin{equation}
        \tau=\ln\Theta,
        \qquad
        s=\frac{1}{\alpha}\bigl(\ln\Psi_\alpha-(1-2\alpha)\tau\bigr),
        \qquad
        \rho=e^s,
    \end{equation}
    \begin{align}
        \psi_j&=k\ln\vartheta_j,
        &
        A_j&=\ell\ln\mathcal M_j,
        \nonumber\\
        \bm{u}&=-2\mu_c\bm{\nabla}\tau+\bm{\nabla}\times\bm{\psi},
        &
        \bm{B}&=\bm{\nabla}\times\bm{A}.
    \end{align}
    The density reconstruction requires \(\alpha\neq0\).
    The incompressible limit \(s=0\), \(\tau=0\), and \(\rho=1\) reduces \(\bm{K}\) to \(-\bm{u}\times\bm{\omega}-\bm{J}\times\bm{B}\), so \eqref{eq:mhd_psi_eq_app} and \eqref{eq:mhd_A_eq_app} recover the velocity- and magnetic-potential structure used by Vanon and Ohkitani.

\section{Restriction on a pure density Schr\"odinger reduction}
\label{app:obstruction}
    It is natural to ask whether the pure density amplitude
    \begin{equation}
        R_\alpha \equiv \rho^\alpha
    \end{equation}
    can itself satisfy a closed Schr\"odinger-type equation. From the continuity equation,
    \begin{equation}
        \partial_t R_\alpha + \bm{u}\cdot\bm{\nabla} R_\alpha
        =
        2\alpha\mu_c (\Delta\tau) R_\alpha.
        \label{eq:Ralpha_transport}
    \end{equation}
    Using \eqref{eq:u_tau_chi}, this becomes
    \begin{equation}
        \partial_t R_\alpha
        +
        2\mu_s \bm{\nabla}^\perp\chi \cdot \bm{\nabla} R_\alpha
        =
        2\mu_c \bm{\nabla}\tau\cdot\bm{\nabla} R_\alpha
        +
        2\alpha\mu_c (\Delta\tau) R_\alpha.
        \label{eq:Ralpha_obstruction}
    \end{equation}
    Thus a first-order compressive transport term remains. More generally, for any smooth local point transform of the density alone,
    \begin{equation}
        R_g \equiv e^{g(s)} = F(\rho),
    \end{equation}
    the continuity equation gives
    \begin{equation}
        \partial_t R_g + \bm{u}\cdot\bm{\nabla} R_g
        =
        2\mu_c g'(s) (\Delta\tau) R_g.
        \label{eq:Rg_transport}
    \end{equation}
    This evolution generates no Laplacian of \(R_g\): every density-only point transform remains a first-order transport-reaction equation. Consequently, a closed Schr\"odinger-type equation for a pure density variable cannot arise from a pointwise reparametrization of \(\rho\) alone.

    We then enlarge the ansatz to a local mixed transform
    \begin{equation}
        \Psi_{g,\beta} \equiv e^{g(s)+\beta\tau}.
    \end{equation}
    A direct calculation gives
    \begin{equation}
        (\partial_t+\bm{u}\cdot\bm{\nabla})\Psi_{g,\beta}
        =
        D \Delta\Psi_{g,\beta}
        +
        U_{g,\beta;D}\Psi_{g,\beta},
    \end{equation}
    with
    \begin{align}
        U_{g,\beta;D}
        &=
        \mu_c\bigl(2g'(s)+\beta\bigr)\Delta\tau
        + \mu_c \beta |\bm{\nabla}\tau|^2
        \nonumber\\
        &\quad
        + \beta\bigl(V_\Theta+\bm{u}\cdot\bm{\nabla}\tau\bigr)
        - D\Bigl(
        g'(s)\Delta s + \beta\Delta\tau
        \nonumber\\
        &\quad
        + \bigl(g''(s)+g'(s)^2\bigr)|\bm{\nabla} s|^2
        + \beta^2 |\bm{\nabla}\tau|^2
        \nonumber\\
        &\quad
        + 2\beta g'(s)\bm{\nabla} s\cdot\bm{\nabla}\tau
        \Bigr).
        \label{eq:Ugbeta}
    \end{align}
    Requiring a constant diffusion coefficient \(D\) and cancellation of the explicit \(\Delta\tau\) and \(|\bm{\nabla}\tau|^2\) terms gives
    \begin{equation}
        D\beta = \mu_c,
        \qquad
        2g'(s)+\beta = 1.
        \label{eq:gbeta_constraints}
    \end{equation}
    Hence \(g'(s)\) must be constant, so \(g(s)=\alpha s+\mathrm{const}\), and therefore
    \begin{equation}
        \beta = 1-2\alpha.
    \end{equation}
    Thus the family \(\Psi_\alpha=\rho^\alpha\Theta^{1-2\alpha}\) already exhausts the local separable transforms of the form \(e^{g(s)+\beta\tau}\) with constant diffusion and exact elimination of the \(\tau\)-Laplacian terms.

\section{Remarks on other transform classes}
\label{app:other_transforms}
    We also tested broader transform classes to check whether the density restriction can be avoided.

\subsection*{1. Local transforms involving gradients}
    First consider a scalar transform that depends locally on the fields and their first spatial
    derivatives,
    \begin{equation}
        Q = F\bigl(s,\tau,\chi,\bm{\nabla} s,\bm{\nabla}\tau,\bm{\nabla}\chi\bigr).
    \end{equation}
    The key differentiated variables obey
    \begin{align}
        (\partial_t+\bm{u}\cdot\bm{\nabla})\bm{\nabla} s
        &=
        2\mu_c \bm{\nabla}\Delta\tau - (\bm{\nabla}\bm{u})^{\!\top}\bm{\nabla} s,
        \label{eq:Dt_grads}
        \\
        \partial_t \bm{\nabla}\tau
        &=
        \mu_c \bm{\nabla}\Delta\tau + \bm{\nabla}\!\bigl(\mu_c |\bm{\nabla}\tau|^2 + V_\Theta\bigr),
        \label{eq:Dt_gradtau}
        \\
        \partial_t \bm{\nabla}\chi
        &=
        \nu \bm{\nabla}\Delta\chi - \frac{1}{2\mu_s}\bm{\nabla}\Psi_F.
        \label{eq:Dt_gradchi}
    \end{align}
    Hence any nontrivial dependence on \(\bm{\nabla} s\), \(\bm{\nabla}\tau\), or \(\bm{\nabla}\chi\) introduces third-order derivatives such as \(\bm{\nabla}\Delta\tau\) or \(\bm{\nabla}\Delta\chi\). Therefore this local \emph{scalar} gradient transform does not close to a second-order Schr\"odinger-type equation unless the gradient dependence is absent.

\subsection*{2. Fixed spatial nonlocal transforms}
    Let \(K\) be a time-independent spatial operator, for example \(\Delta^{-1}\), \(\bm{\nabla}\Delta^{-1}\), or \((-\Delta)^{-1/2}\), and define
    \begin{equation}
        Q_K \equiv K R_g,
        \qquad
        R_g = F(\rho).
    \end{equation}
    Using \eqref{eq:Rg_transport}, we find
    \begin{equation}
        \partial_t Q_K + \bm{u}\cdot\bm{\nabla} Q_K
        =
        [\bm{u}\cdot\bm{\nabla}, K] R_g
        +
        2\mu_c K\!\bigl(g'(s)\Delta\tau\,R_g\bigr).
        \label{eq:nonlocal_commutator}
    \end{equation}
    The transformed equation acquires the nonlocal commutator \([\bm{u}\cdot\bm{\nabla}, K]\), which vanishes only in very special flows. Thus a fixed nonlocal spatial transform does not naturally produce a local Schr\"odinger-type operator.

\subsection*{3. Auxiliary-field and doubled-amplitude formulations}
    Finally, we can write an exact doubled Schr\"odinger-type formulation. For any \(\alpha\neq \frac12\), we define the partner amplitudes
    \begin{equation}
        \Psi_\alpha = \rho^\alpha \Theta^{1-2\alpha},
        \qquad
        \Psi_{1-\alpha} = \rho^{1-\alpha}\Theta^{2\alpha-1}.
    \end{equation}
    Their product recovers the density exactly:
    \begin{equation}
        \rho = \Psi_\alpha \Psi_{1-\alpha}.
        \label{eq:rho_partner_product}
    \end{equation}
    Moreover, each partner satisfies an equation of the form \eqref{eq:Psi_alpha_vector},
    with parameters
    \begin{equation}
        D_{1-\alpha} = - D_\alpha,
        \qquad
        \bm{\mathcal A}_{1-\alpha} = - \bm{\mathcal A}_\alpha.
        \label{eq:partner_symmetry}
    \end{equation}
    Thus the continuity sector admits an exact doubled vector-potential-coupled Schr\"odinger representation, although not a single pure density-carrying one.

\bibliography{bib}

\end{document}